\documentclass[a4paper]{article}

\usepackage[top=25mm, bottom=25mm, left=25mm, right=25mm]{geometry}
\input{Qcircuit}

\usepackage{algorithmicx}
\usepackage[ruled]{algorithm}
\usepackage[noset]{algpseudocode}

\newcommand{\euk}{}
\newenvironment{alginc}[1][pseudocode]{\medskip\algsetlanguage{#1}\begin{algorithmic}[1]}{\end{algorithmic}\medskip}

\newcommand\ASTART{\bigskip\noindent\begin{minipage}[c]{0.5\linewidth}}

\newcommand\AENDSKIP{\end{minipage}\bigskip}
\newcommand\AEND{\end{minipage}}

\newtheorem{theorem}{Theorem}[section]

\newtheorem{definition}[theorem]{Definition}

\newcommand{\qed}{\nobreak \ifvmode \relax \else
      \ifdim\lastskip<1.5em \hskip-\lastskip
      \hskip1.5em plus0em minus0.5em \fi \nobreak
      \vrule height0.75em width0.5em depth0.25em\fi}

\title{Tight Bounds on the Synthesis of 3-bit Reversible Circuits: NFT Library}

\author{Ahmed Younes\footnote {ayounes2@yahoo.com or ayounes@alexu.edu.eg}\\
Department of Mathematics and Computer Science\\
Faculty of Science, Alexandria University\\
Alexandria, Egypt}

\begin{document}
\maketitle
\begin{abstract}

The reversible circuit synthesis problem can be reduced to permutation group. This allows  
Schreier-Sims Algorithm for the strong generating set-finding problem to be used to find tight bounds 
on the synthesis of 3-bit reversible circuits using the NFT library. The tight bounds include 
the maximum and minimum length of 3-bit reversible circuits, 
the maximum and minimum cost of 3-bit reversible circuits. The analysis shows better results than 
that found in the literature for the lower bound of the cost. The analysis also shows that there are 
1960 universal reversible sub-libraries from the main NFT library.


\end{abstract}

\section{Introduction}  

Reversible logic \cite{bennett73,fredtoff82} is one of the hot areas of research. It has many applications 
in quantum computation \cite{Gruska99,nc00a}, low-power CMOS \cite{cmos2,cmos1} and many more. 
Synthesis of reversible circuits cannot be done using conventional ways \cite{toffoli80}. 
Synthesis and optimization of Boolean systems on non-standard computers that promise to do computation
more powerfully \cite{simon94} than classical computers, such as
quantum computers, is an essential aim in the exploration of the
benefits that may be gain from such systems.

A lot of work has been done trying to find an efficient reversible circuit for an arbitrary reversible 
function. In one of the research 
directions, it was shown that the process of synthesizing linear reversible circuits can be reduced to 
a row reduction problem of $n \times n$ non-singular matrix \cite{patel}. Standard row reduction methods such as Gaussian 
elimination and LU-decomposition have been proposed \cite{Beth01}. In another research direction, 
search algorithms and template matching tools using reversible gates libraries have been used 
\cite{Dueck,Maslov1,DMMiller2,DMMiller1}. These will work efficiently for small circuits. 
A method is given in \cite{transrules}, where a very useful set
of transformations for Boolean quantum circuits is shown. In this method, extra
auxiliary bits are used in the construction that will increase the hardware cost. In \cite{Younes03b}, 
it was shown that there is a direct correspondence between reversible 
Boolean operations and certain forms of classical logic known as Reed-Muller expansions. 
This arises the possibility of handling the problem of synthesis and optimization of reversible Boolean
logic within the field of Reed-Muller logic. A lot of work has been done trying to find an efficient 
reversible circuit for an arbitrary multi-output Boolean functions by using templates 
\cite{Maslov2,template} and data-structure-based optimization \cite{datastr}. 
A method to generate an optimal 4-bit reversible circuits has been proposed \cite{optimal4}. 
Benchmarks for reversible circuits have been established \cite{Benurl2}.

Recently, the study of reversible logic synthesis problem using group theory 
is gaining more attention. 
Investigation on the universality of the basic building blocks of reversible circuit has been done \cite{group1,group2}. 
A relation between Young subgroups and the reversible logic synthesis problem has been proposed \cite{group3}. 
A comparison between the decomposition of reversible circuit and quantum circuit using group theory has been shown \cite{group4}. 
A GAP-based algorithms to synthesize reversible circuits for various types of gate with various gate costs has been proposed \cite{art}.

The aim of the paper is to answer the following questions for the synthesis of 3-bit reversible circuits 
using NFT library:

\begin{enumerate}

\item What are the maximum and minimum lengths of reversible circuits for 
a reversible function? What are the costs of these circuits?

\item What are the maximum and minimum costs of reversible circuits for 
a reversible function ? What are the length of these circuits?

\item What are the upper-bound and lower-bound on lengths of reversible circuits 
for all 3-bit reversible functions using NFT library? What are the costs of these bounds?
 
\item What are the upper-bound and lower-bound on costs of reversible circuits 
for all 3-bit reversible functions using NFT library? What are the lengths of these bounds?

\item Is there any sub-library of $NFT$ that can act as a universal reversible gate library? If so, 
what is the best universal reversible gate library that gives the best length, worst length, 
best cost and worst cost? How many sub-libraries that can act as a universal reversible gate library?

\end{enumerate}

The paper is organized as follows: 
Section 2 gives a short background on the synthesis of reversible circuit 
problem and shows the reduction the problem to permutation group. 
Section 3 shows Schreier-Sims Algorithm for the strong generating set-finding problem 
that is used to calculate the bounds on the synthesis of 3-bit reversible circuits 
problem using NFT library. 
Section 4 shows the results of the experiments. The paper ends up
with a summary and conclusion in Section 5.

\section{Background}
\label{sec2}

\begin{center}
\begin{figure}[htbp]
\begin{center}
\setlength{\unitlength}{3947sp}%
\begingroup\makeatletter\ifx\SetFigFont\undefined%
\gdef\SetFigFont#1#2#3#4#5{%
  \reset@font\fontsize{#1}{#2pt}%
  \fontfamily{#3}\fontseries{#4}\fontshape{#5}%
  \selectfont}%
\fi\endgroup%
\begin{picture}(3169,1438)(5451,-1937)
{\thinlines
\put(6301,-586){\circle*{150}}
}%
{\put(6301,-811){\circle*{150}}}%
{\put(6301,-1261){\circle*{150}}}%
{\put(6301,-1561){\circle{150}}}%
{\put(7658,-1404){\circle{150}}}%
{\put(7661,-1231){\circle*{150}}}%
{\put(8376,-1322){\circle{150}}}%
{\put(8373,-1136){\circle*{150}}}%
{\put(8372,-949){\circle*{150}}}%
{\put(7645,-649){\circle{150}}}%
{\put(6751,-811){\line(-1, 0){900}}}%
{\put(6751,-1261){\line(-1, 0){900}}}%
{\put(6751,-1561){\line(-1, 0){900}}}%
{\put(6301,-511){\line( 0,-1){450}}}%
\put(6270,-1100){\makebox(0,0)[lb]{{$\vdots$}}}

{\put(6301,-1111){\line( 0,-1){525}}}%
{\put(6751,-586){\line(-1, 0){900}}}%
{\put(7424,-1407){\line( 1, 0){463}}}%
{\put(7659,-1162){\line( 0,-1){318}}}%
{\put(7422,-1233){\line( 1, 0){463}}}%
{\put(8145,-1325){\line( 1, 0){463}}}%
{\put(8372,-877){\line( 0,-1){520}}}%
{\put(8137,-1142){\line( 1, 0){463}}}%
{\put(8140,-959){\line( 1, 0){463}}}%
{\put(7418,-650){\line( 1, 0){463}}}%
{\put(7644,-582){\line( 0,-1){140}}}%

\put(5476,-597){\makebox(0,0)[lb]{{$x_{1}$}}}
\put(5476,-814){\makebox(0,0)[lb]{{$x_{2}$}}}
\put(5476,-1267){\makebox(0,0)[lb]{{$x_{n-1}$}}}
\put(5476,-1561){\makebox(0,0)[lb]{{$f_{in}$}}}

\put(6800,-1561){\makebox(0,0)[lb]{{$f_{out}$}}}
\put(6800,-1267){\makebox(0,0)[lb]{{$y_{n-1}$}}}
\put(6800,-814){\makebox(0,0)[lb]{{$y_{2}$}}}
\put(6800,-597){\makebox(0,0)[lb]{{$y_{1}$}}}

\put(7130,-1950){\makebox(0,0)[lb]{{c.Feynman}}}
\put(8131,-1922){\makebox(0,0)[lb]{{d.Toffoli}}}
\put(7350,-939){\makebox(0,0)[lb]{{b.NOT}}}
\put(6000,-1922){\makebox(0,0)[lb]{{a.$C^{n}NOT$}}}
\end{picture}%

\end{center}
\caption{$C^{n}NOT$ gates. The black circle $\bullet $
indicates the control bits, and the symbol $ \oplus $ indicates
the target bit. (a)$C^{n}NOT$ gate with $n-1$ control bits. (b) $C^{1}NOT$
gate with no control bits. (c)$C^{2}NOT$ gate with one control bit.
(d)$C^{3}NOT$ gate with two control bits.}
\label{figcnot}
\end{figure}
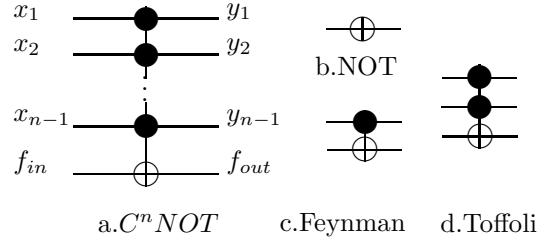
\end{center}

\subsection{Reversible Circuits}

To build a reversible circuit with $n$ variables, an $n \times n$ reversible
circuit is used. $C^{n}NOT$ gate is the main primitive gate
that is used to build the circuit since it is shown to be universal for 
reversible computation \cite{toffoli80}. $C^{n}NOT$ gate is defined
as follows:

\begin{definition}($C^{n}NOT$ gate)

$C^{n}NOT$ is a reversible gate denoted as,
\begin{equation} 
C^{n}NOT(x_{1}, x_{2},\ldots,x_{n-1};f),
\end{equation}
\noindent
with
$n$ inputs: $x_{1}$, $x_{2}$,$\ldots,x_{n-1}$ (known as control
bits) and $f_{in}$ (known as target bit), and $n$ outputs:
$y_{1}$, $y_{2}$,$\ldots,y_{n-1}$ and ${f_{out}}$. The operation
of the $C^{n}NOT$ gate is defined as follows,

\begin{equation}
\begin{array}{l}
 y_i  = x_i ,\mathrm{for}\,1 \le i \le n-1, \\
 f_{out}  = f_{in}  \oplus x_{1} x_{2}  \ldots x_{n-1}, \\
 \end{array}
\end{equation}
\noindent
\end{definition}
\noindent
i.e. the target bit will be flipped if and only if all the control
bits are set to 1. Some special cases of the general $C^{n}NOT$ gate
have their own names, $C^{1}NOT$ gate with no control bits is called
$NOT$ gate as shown in Fig.~\ref{figcnot}-b, where the bit will be
flipped unconditionally. $C^{2}NOT$ gate with one control bit is
called {\it Feynman} gate as shown in Fig.~\ref{figcnot}-c. $C^{3}NOT$ gate
with two control bits is called {\it Toffoli} gate as shown in 
Fig.~\ref{figcnot}-d.  

For the sake of readability  
$C^{1}NOT$, $C^{2}NOT$ and $C^{3}NOT$ will be written 
for short as $N$, $F$ and $T$ respectively where the control and/or target 
bits will be shown in the subscript of the gate, for example, a 3-bit reversible circuit 
synthesized using $N$,$F$ and $T$ gates can contain 12 possible gates, 
as shown in Fig.\ref{12gates}, that perform as follows:

\begin{figure} [htbp]
\begin{center}
\[
\Qcircuit @C=0.7em @R=0.5em @!R{
\lstick{x_1}	&	&\targ	&\qw		&\qw		&\qw	&\qw		&\qw	&\ctrl{1}	&\qw	&\ctrl{2}	&\qw	&\qw		&\qw	&\targ		&\qw	&\qw		&\qw	&\targ		&\qw	&\ctrl{2}	&\qw	&\ctrl{1}	&\qw	&\targ			&\qw		\\
\lstick{x_2}	&	&\qw	&\qw		&\targ		&\qw	&\qw		&\qw	&\targ		&\qw	&\qw		&\qw	&\ctrl{1}	&\qw	&\ctrl{-1}	&\qw	&\targ		&\qw	&\qw		&\qw	&\ctrl{1}	&\qw	&\targ		&\qw	&\ctrl{-1}		&\qw		\\
\lstick{x_3}	&	&\qw	&\qw		&\qw		&\qw	&\targ		&\qw	&\qw		&\qw	&\targ		&\qw	&\targ		&\qw	&\qw		&\qw	&\ctrl{-1}	&\qw	&\ctrl{-2}	&\qw	&\targ		&\qw	&\ctrl{-1}	&\qw	&\ctrl{-2}		&\qw		\\
			&	&N_1	&			&N_2		&		&N_3		&		&F_{12}		&		&F_{13}		&		&F_{23}		&		&F_{21}		&		&F_{32}		&		&F_{31}		&		&T_{123}	&		&T_{132}	&		&T_{321}		&		
}
\]
\end{center}
\caption{The 12 possible CNOT gates for a 3 bit reversible circuit.}
\label{12gates}
\end{figure}
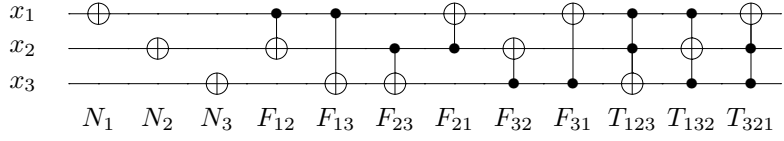

\begin{equation}
\begin{array}{l}
N_1:(x_1,x_2,x_3)\to (x_1\oplus1,x_2,x_3)\\
N_2:(x_1,x_2,x_3)\to (x_1,x_2\oplus1,x_3)\\
N_3:(x_1,x_2,x_3)\to (x_1,x_2,x_3\oplus1)\\
F_{12}:(x_1,x_2,x_3)\to (x_1,x_2\oplus x_1,x_3)\\
F_{13}:(x_1,x_2,x_3)\to (x_1,x_2,x_3\oplus x_1)\\
F_{23}:(x_1,x_2,x_3)\to (x_1,x_2,x_3\oplus x_2)\\
F_{21}:(x_1,x_2,x_3)\to (x_1\oplus x_2,x_2,x_3)\\
F_{32}:(x_1,x_2,x_3)\to (x_1,x_2\oplus x_3,x_3)\\
F_{31}:(x_1,x_2,x_3)\to (x_1\oplus x_3,x_2,x_3)\\
T_{123}:(x_1,x_2,x_3)\to (x_1,x_2,x_3\oplus x_1 x_2)\\
T_{132}:(x_1,x_2,x_3)\to (x_1,x_2\oplus x_1 x_3,x_3)\\
T_{321}:(x_1,x_2,x_3)\to (x_1\oplus x_2 x_3,x_2,x_3)
\end{array}
\label{12gatelogic}
\end{equation}		

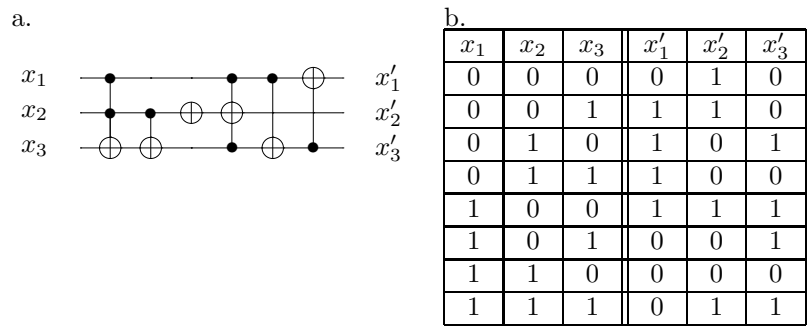
\begin{figure} [htbp]
\begin{center}
\begin{tabular}
{p{150pt}p{100pt}}
a.
\[
\Qcircuit @C=0.7em @R=0.5em @!R{
\lstick{x_1}&	&\ctrl{2}		&\qw			&\qw			&\ctrl{1}		&\ctrl{2}		&\targ		&\qw		&\rstick{x_1'}\\
\lstick{x_2}&	&\ctrl{1}		&\ctrl{1}		&\targ		&\targ		&\qw			&\qw			&\qw		&\rstick{x_2'}\\
\lstick{x_3}&	&\targ		&\targ		&\qw			&\ctrl{-1}		&\targ		&\ctrl{-2}		&\qw		&\rstick{x_3'}
}
\]
& 
b.
\begin{tabular}{|c|c|c||c|c|c|}
\hline $x_1$  & $x_2$& $x_3$   & $x_1'$  & $x_2'$& $x_3'$  \\
\hline
0&0&0&0&1&0  \\ \hline
0&0&1&1&1&0 \\ \hline
0&1&0&1&0&1 \\ \hline
0&1&1&1&0&0 \\ \hline
1&0&0&1&1&1  \\ \hline
1&0&1&0&0&1  \\ \hline
1&1&0&0&0&0  \\ \hline
1&1&1&0&1&1 \\ \hline
\end{tabular}%
\\
\end{tabular}
\end{center}
\caption{(a.) Circuit implementation of a reversible function. (b.) Reversible truth table. }
\label{revcirTT}
\end{figure}

A reversible circuit is a cascade of one or more $C^{n}NOT$ gates, 
for example, a 3-bit reversible circuit shown in Fig.\ref{revcirTT}-a can be written for short 
as $(T_{123}F_{23}N_2T_{132}F_{13}F_{31})$, 
where the truth table of this circuit is shown in Fig.\ref{revcirTT}-b. Another way to write 
this truth table is $(0,1,2,3,4,5,6,7)\to(2,6,5,4,7,1,0,3)$, or simply $(2,6,5,4,7,1,0,3)$ which 
is called the specification of the circuit. 	

\subsection{Quantum Cost}

{\it Quantum cost} is a term used to refer to
the technological cost of building $C^{n}NOT$ gates. The quantum
cost of a reversible circuit is subject to
optimization as well as the number of $C^{n}NOT$ gates used in the
circuit.  The quantum cost of a $C^{n}NOT$ gate is based primarily on 
the number of bits involved in the gate, i.e. the number of
elementary operations required to build the $C^{n}NOT$ gate
\cite{PhysRevA.52.3457}. 

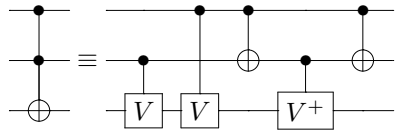
\begin{figure} [htbp]
\begin{center}
\[
\Qcircuit @C=0.7em @R=0.5em @!R{
&\ctrl{2}	&\qw	&			&	&\qw				&\ctrl{1}			&\ctrl{1}		&\qw				&\ctrl{1}		&\qw		&\\
&\ctrl{1}	&\qw	& \equiv	&	&\ctrl{1}			&\qw\qwx[1]		&\targ		&\ctrl{1}			&\targ		&\qw					&\\
&\targ		&\qw	& 			&	&\gate{V}			&\gate{V}			&\qw			&\gate{V^{+}}		&\qw			&\qw		&
}
\]
\end{center}
\caption{Decomposition of $T$ gate as 5 elementary 2-qubit gates.}
\label{toffdeca}
\end{figure}

When implementing a reversible gate circuit on a quantum computer,
four elementary quantum gates are used \cite{PhysRevA.52.3457}: 
$N$ gate, $F$ gate, Controlled-$V$ and Controlled-$V^{+}$ gates, 
where $V · V = V^{+}·V^{+} = N$, $V · V^{+} =V^{+} · V = I$, 
where $I$ is the identity. The cost of a 2-qubit gate is much larger than 
that of a 1-qubit gate. In this paper, the cost of
$N$ gate is ignored as in \cite{art} to be able to compare results, 
and assume that the cost of any 2-qubit gate is equal to 1. 
The cost of a reversible circuit is measured by the number of 2-qubit gates used in its 
implementation as a quantum circuit, so, the quantum cost of $T$ gate is 5 as shown in 
Fig.\ref{toffdeca}. 

\subsection{Basic Notions}

In this section, the basic notions for reversible circuit, permutation group, NFT library, 
the relationship between reversible logic circuits and permutation group theory will be reviewed.

\begin{definition}
Let $X = \{0, 1\}$. A Boolean function $f$ with $n$ input variables,
$x_1, \ldots, x_n$, and $n$ output variables, $y_1,\ldots, y_n$, is a function 
$f : X^n \to X^n$, where $(x_1, \ldots, x_n) \in X^n$ 
is called the input vector and $(y_1, \ldots, y_n) \in X^n$ 
is called the output vector.
\end{definition}

\begin{definition}
An $n$-input $n$-output Boolean function is reversible  ($n \times n$ function) 
if it maps each input vector to a unique output vector, i.e. a one-to-one, 
onto function (bijection). 
\end{definition}

There are $2^n!$ reversible $n \times n$ Boolean functions. For $n$ = 3, 
there are 40,320 $3$-input $3$-output reversible functions.

\begin{definition}

An $n$-input $n$-output reversible gate (or circuit) is a gate 
 that realizes an $n \times n$ reversible function.
\end{definition}

\begin{definition}
 
A set of reversible gates that can be used to
build a reversible circuit is called a gate library $L$.

\end{definition}

\begin{definition}
 
A universal reversible gate library $L_n$ is a set of reversible gates that can be used to
build any reversible circuit with $n$-input $n$-output.

\end{definition}

\begin{definition}

Consider a finite set $A=\{1,2,...,N\}$ and a bijection $\sigma :A \to A$,
then $\sigma$ can be written as,

\begin{equation}
\sigma  = \left( {\begin{array}{*{20}c}
   1 & 2 & 3 & {...} & {N}  \\
   {\sigma (1)} & {\sigma (2)} & {\sigma (3)} & {...} & {\sigma (N)}  \\
\end{array}} \right),
\end{equation}

\noindent i.e. $\sigma$ is a permutation of $A$. Let $A$ be an ordered set, 
then the top row can be eliminated and $\sigma$ can be written as,

\begin{equation}
\left( {\sigma (1)} , {\sigma (2)} , {\sigma (3)} , {...} , {\sigma (N)}\right).
\label{specseqn}
\end{equation}

Any reversible circuit with $n$-input $n$-output can be considered as 
a permutation $\sigma$ and Eqn.\ref{specseqn} is called the specification 
of this reversible circuit such that $N=2^n$.
\end{definition}

The set of all permutations on $A$ forms a symmetric group on $A$ under 
composition of mappings \cite{Ref6}, denoted by $S_N$ \cite{Ref2}. 
A permutation group $G$ is a subgroup \cite{Ref6} of the
symmetric group $S_N$. A universal reversible gate library $L_n$ is called the generators of 
the group. Another important notation of a permutation is the product of disjoint cycles \cite{Ref2}. 
For example,
$\left( {\begin{array}{*{20}c}
   {1,2,3,4,5,6,7,8}  \\
   {3,2,5,4,6,1,8,7}  \\
\end{array}} \right)$
will be written as (1, 3, 5, 6)(7, 8). The identity mapping
"()" is called the unit element in a permutation group.
A product $p * q$ of two permutations $p$ and $q$ means applying
mapping $p$ then $q$, which is equivalent to cascading $p$ and $q$. 

The one-to-one correspondence between a $n \times n$ reversible circuit
and a permutation on $A = \{1, 2, . . ., N\}$ is established as done in \cite{art}. 
In the permutation group references, $A$ begins from one, instead of zero. 
Therefore, we have the following relation \cite{art}:
$<X_N, X_{N-1},\ldots, X_{1} >_2= index(X_N, \ldots, X_1) - 1$. 
Using the integer coding, a permutation is considered as a bijective function $f : \{1, 2, \ldots, N\} \to
\{1, 2, \ldots, N\}$. Cascading two generators is equivalent to multiplying two permutations.
In what follows, a $n \times n$ reversible gate is not distinguished  from a
permutation in $S_{N}$. The set of 12 gates shown in Fig. \ref{12gates} and performs as shown 
in Eqn.\ref{12gatelogic} can be re-written as product of disjoint cycles as follows,

\begin{equation}
\begin{array}{l}
N_1:(1,5)(2,6)(3,7)(4,8)\\
N_2:(1,3)(2,4)(5,7)(6,8)\\
N_3:(1,2)(3,4)(5,6)(7,8)\\
F_{12}:(5,7)(6,8)\\
F_{13}:(5,6)(7,8)\\
F_{23}:(3,4)(7,8)\\
F_{21}:(3,7)(4,8)\\
F_{32}:(2,4)(6,8)\\
F_{31}:(2,6)(4,8)\\
T_{123}:(7,8)\\
T_{132}:(6,8)\\
T_{321}:(4,8)
\end{array}
\label{12gatecycle}
\end{equation}

\section{Schreier-Sims Algorithm}

Let $A$ be a finite set. The Symmetric group, $Sym(A)$, is the group of all bijections 
from $A$ to itself. A permutation group $G$ is a subgroup of $Sym(A)$. 
Let $G\le Sym(A)$. For $\alpha \in A$, let $G_{\alpha}$ denotes 
the stabilizer of $\alpha$ in $G$, i.e., $G_\alpha   = \{ g \in G|\alpha ^g  = \alpha$. 
Let $B=(\alpha_1,\alpha_2,\ldots,\alpha_k)$ with $\alpha_i \in G$, then $B$ is a base for $G$ 
if $G_{(\alpha_1,\alpha_2,\ldots,\alpha_k)}={1}$, and the chain of subgroups,

\begin{equation}
G = G^{(1)}  \ge G^{(2)}  \ge  \ldots  \ge G^{(k + 1)}  = \{ 1\},
\end{equation}

\noindent defined by $G^{(i+1)}=G^{(i)}_{\alpha_i}$ for 
$1\le i \le k$ is called the stablizer chain for $B$. The orbit of $\alpha$ under $G$, denoted 
$\alpha^G$ is the set $\alpha^G=\{\alpha^g|g \in G \}$. A strong generating set for $G$ relative to $B$, denoted $S$, is a set $S \subseteq G$ and for every $i$ with 
$1 \le i \le k+1$, $G^{(i)}=\left\langle {S \cap G^{(i)} } \right\rangle$ holds.

This allows the problem of synthesizing a minimal number of gates (generators) to be reduced 
to a strong generating set-finding problem, that is, given a group $G$ acting 
faithfully on a finite set $A$ of size $N$. $G$ is specified by means of a generating 
set $S$ where each element of $S$ is expressed as a permutation on $A$.

Schreier-Sims Algorithm is a poly-time algorithm \cite{algoref} known in computational Group Theory that solves 
the strong generating set-finding problem and also 
go further to solve membership testing problem. 
Schreier-Sims Algorithm keeps computing the generating sets using cosets, 
by trimming down the size of the generating set at every step using 
a depth first search approach to keep the size of the generating set from growing 
too large. An implementation of Schreier-Sims Algorithm on GAP \cite{GAP2013} has been used 
to find the minimal number of generators that generates a specification of 
a reversible circuit given that the generators of the group can represent that specification .

Given the universal gate library NFT with 12 generators as shown in Eqn.\ref{12gatecycle} 
and the 40320 specifications for all 3-input 3-output reversible circuits. The aim of the 
experiment is to answer the questions shown in the aim of the paper. 
To answer these questions, all sub-libraries of NFT library has been generated, that is 4095 sub-libraries 
after excluding the identity mapping. 
Use every sub-library to try to  synthesize a reversible circuit for the 40320 specifications, 
if possible, using Schreier-Sims Algorithm. The term "if possible" here means that if a specification 
does not belong to the group generated by a sub-library, then it is impossible for 
this specification to be represented as a reversible circuit using this sub-library. 
The process of synthesizing all possible 3-bit reversible circuits is shown in Algorithim \ref{EIdent}.

\begin{algorithm}[H]
\caption{Generation of All 3 bits reversible circuits}\label{EIdent}
\begin{alginc}
\Procedure{\euk}{$ $}
	\For{$i\gets 1, NumOfSubLibraries$}
			\For{$j\gets 1, NumOfCircuits$}
				\If{$Specs[j] \in G[i]$}
					\State $Circuit[i][j]=SchreierSims(G[i],Specs[j])$
				\Else
					\State $Circuit[i][j]=[ ]$
				\EndIf
            \EndFor	
     \EndFor
\EndProcedure
\end{alginc}
\end{algorithm}

\section{Expermimental Results}

Among the $4,095 \times 40,319 =165,106,305$ trials to synthesize a circuit for a specification using a sub-library, 
80,925,627 circuits have been synthesized. The trivial 
specification "()" and the trivial identity mapping have been ignored.
The results have been organized to answer the specification related questions (Question 1 to Question 4) in
Section \ref{sor}, and the question related to the sub-libraries (Question 5) has been answered 
in Section \ref{lor}.

\subsection{Specification Oriented Results}
\label{sor}

In this section, the results are organized to show a summary for the specification oriented 
results in Table \ref{tab1}. The results related 
to the bounds on circuits length are shown in Tables \ref{tab2}, \ref{tab3}, 
\ref{tab4} and \ref{tab5}. Then the results related to the bounds on circuits 
cost are shown in Tables \ref{tab6}, \ref{tab7}, \ref{tab8} and \ref{tab9}.

\begin{table}[htbp]
\small
\begin{center}
\begin{tabular}
{|p{48pt}|p{48pt}|p{1pt}|p{48pt}|p{48pt}|p{1pt}|p{48pt}|p{48pt}|}
\cline{1-2}
\cline{4-5}
\cline{7-8}
{\#}lib& 
{\#}specs& 
& 
{\#}lib& 
{\#}specs& 
& 
{\#}lib& 
{\#}specs \\
\cline{1-2}
\cline{4-5}
\cline{7-8}
1960& 
29670& 
& 
2263& 
108& 
& 
2468& 
12 \\
\cline{1-2}
\cline{4-5}
\cline{7-8}
1984& 
216& 
& 
2264& 
24& 
& 
2496& 
6 \\
\cline{1-2}
\cline{4-5}
\cline{7-8}
2016& 
1746& 
& 
2266& 
80& 
& 
2525& 
24 \\
\cline{1-2}
\cline{4-5}
\cline{7-8}
2044& 
36& 
& 
2268& 
12& 
& 
2528& 
27 \\
\cline{1-2}
\cline{4-5}
\cline{7-8}
2080& 
1170& 
& 
2274& 
12& 
& 
2540& 
6 \\
\cline{1-2}
\cline{4-5}
\cline{7-8}
2085& 
1& 
& 
2284& 
48& 
& 
2560& 
27 \\
\cline{1-2}
\cline{4-5}
\cline{7-8}
2086& 
559& 
& 
2287& 
135& 
& 
2605& 
12 \\
\cline{1-2}
\cline{4-5}
\cline{7-8}
2116& 
540& 
& 
2311& 
27& 
& 
2624& 
13 \\
\cline{1-2}
\cline{4-5}
\cline{7-8}
2120& 
96& 
& 
2320& 
210& 
& 
2625& 
8 \\
\cline{1-2}
\cline{4-5}
\cline{7-8}
2122& 
2974& 
& 
2324& 
24& 
& 
2636& 
6 \\
\cline{1-2}
\cline{4-5}
\cline{7-8}
2128& 
162& 
& 
2335& 
9& 
& 
2676& 
12 \\
\cline{1-2}
\cline{4-5}
\cline{7-8}
2132& 
192& 
& 
2348& 
12& 
& 
2688& 
9 \\
\cline{1-2}
\cline{4-5}
\cline{7-8}
2144& 
78& 
& 
2353& 
24& 
& 
2689& 
3 \\
\cline{1-2}
\cline{4-5}
\cline{7-8}
2152& 
36& 
& 
2354& 
12& 
& 
2705& 
3 \\
\cline{1-2}
\cline{4-5}
\cline{7-8}
2176& 
24& 
& 
2393& 
45& 
& 
2732& 
6 \\
\cline{1-2}
\cline{4-5}
\cline{7-8}
2191& 
27& 
& 
2412& 
72& 
& 
2816& 
7 \\
\cline{1-2}
\cline{4-5}
\cline{7-8}
2196& 
96& 
& 
2417& 
27& 
& 
2880& 
6 \\
\cline{1-2}
\cline{4-5}
\cline{7-8}
2198& 
12& 
& 
2428& 
24& 
& 
2944& 
6 \\
\cline{1-2}
\cline{4-5}
\cline{7-8}
2217& 
1293& 
& 
2432& 
11& 
& 
2961& 
3 \\
\cline{1-2}
\cline{4-5}
\cline{7-8}
2218& 
108& 
& 
2450& 
6& 
& 
3072& 
3 \\
\cline{1-2}
\cline{4-5}
\cline{7-8}
2220& 
72& 
& 
2455& 
27& 
& 
3200& 
3 \\
\cline{1-2}
\cline{4-5}
\cline{7-8}
2246& 
12& 
& 
2456& 
3& 
& 
3264& 
6 \\
\cline{1-2}
\cline{4-5}
2249& 
36& 
& 
2465& 
3& 
& 
& 
 \\
\cline{1-2}
\cline{4-5}
\cline{7-8}
\end{tabular}
\caption{There are ${\#}specs$ specifications that can be synthesized by ${\#}lib$ 
sub-libraries.}
\label{tab1}
\end{center}
\end{table}

\begin{figure} [htbp]
\begin{center}
\[
\Qcircuit @C=0.7em @R=0.5em @!R{
\lstick{}&	&\ctrl{2}	&\qw	  &	  	&\qw			&\ctrl{1}		&\qw			&\ctrl{1}		&\qw			&\qw			&\qw			&\ctrl{1}		&\qw			&\ctrl{1}		&\qw		&\rstick{}\\
\lstick{}&	&\qw		&\qw	&		&\targ		&\targ		&\ctrl{1}		&\targ		&\ctrl{1}		&\targ		&\ctrl{1}		&\targ		&\ctrl{1}		&\targ		&\qw		&\rstick{}\\
\lstick{}&	&\targ		&\qw	  &	  	&\qw			&\ctrl{-1}		&\targ		&\ctrl{-1}		&\targ		&\qw			&\targ		&\ctrl{-1}		&\targ		&\ctrl{-1}		&\qw		&\rstick{}
}
\]
\end{center}
\caption{ The worst circuit representation for $F_{31}$.}
\label{F31worst}
\end{figure}
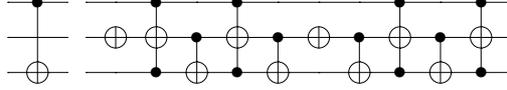

Table \ref{tab1} shows the number of sub-libraries that can be used to synthesize 
a circuit for a specification, for example, there are 1960 sub-libraries that can 
be used to synthesize a circuit for 29670 specifications. 
There are 6 popular specifications that can be synthesized by 3264 different sub-libraries. 
These 6 specifications are the specifications of $F_{12},F_{13},F_{23},F_{21},F_{32}$ and $F_{31}$. 
Fig.\ref{F31worst} shows the worst circuit representation for $F_{31}$ with 10 gates and cost = 24. 
This shows that any $F$ gates can be removed from a library and be replaced 
by another set of generators although there might be an increase in the cost of the circuit.

\begin{table}[htbp]
\small
\begin{center}
\begin{tabular}
{|p{48pt}|p{48pt}|}
\hline
Max Len& 
{\#}specs \\
\hline
7& 
3 \\
\hline
8& 
7 \\
\hline
9& 
16 \\
\hline
10& 
29 \\
\hline
11& 
63 \\
\hline
12& 
154 \\
\hline
13& 
1455 \\
\hline
14& 
14966 \\
\hline
15& 
19544 \\
\hline
16& 
3815 \\
\hline
17& 
243 \\
\hline
18& 
24 \\
\hline
\end{tabular}
\caption{There are $\#specs$ that have circuits with maximum length = $Maxlen$. The average maximum length = 14.639.}
\label{tab2}
\end{center}
\end{table}

\begin{table}[htbp]
\small
\begin{center}
\begin{tabular}
{|p{15pt}|p{15pt}|p{25pt}|p{1pt}|p{15pt}|p{15pt}|p{25pt}|p{1pt}|p{15pt}|p{15pt}|p{25pt}|p{1pt}|p{15pt}|p{15pt}|p{25pt}|}
\cline{1-3}
\cline{5-7}
\cline{9-11}
\cline{13-15}
Max Len& 
cost& 
{\#}specs& 
& 
Max Len& 
cost& 
{\#}specs& 
& 
Max Len& 
cost& 
{\#}specs& 
& 
Max Len& 
cost& 
{\#}specs \\
\cline{1-3}
\cline{5-7}
\cline{9-11}
\cline{13-15}
7& 
9& 
2& 
& 
13& 
19& 
64& 
& 
14& 
39& 
122& 
& 
16& 
22& 
3 \\
\cline{1-3}
\cline{5-7}
\cline{9-11}
\cline{13-15}
7& 
11& 
1& 
& 
13& 
20& 
37& 
& 
14& 
40& 
107& 
& 
16& 
23& 
18 \\
\cline{1-3}
\cline{5-7}
\cline{9-11}
\cline{13-15}
8& 
14& 
3& 
& 
13& 
21& 
40& 
& 
14& 
41& 
109& 
& 
16& 
24& 
251 \\
\cline{1-3}
\cline{5-7}
\cline{9-11}
\cline{13-15}
8& 
20& 
4& 
& 
13& 
22& 
187& 
& 
14& 
42& 
62& 
& 
16& 
25& 
102 \\
\cline{1-3}
\cline{5-7}
\cline{9-11}
\cline{13-15}
9& 
11& 
1& 
& 
13& 
23& 
213& 
& 
14& 
43& 
46& 
& 
16& 
26& 
6 \\
\cline{1-3}
\cline{5-7}
\cline{9-11}
\cline{13-15}
9& 
13& 
2& 
& 
13& 
24& 
72& 
& 
14& 
44& 
52& 
& 
16& 
27& 
66 \\
\cline{1-3}
\cline{5-7}
\cline{9-11}
\cline{13-15}
9& 
17& 
3& 
& 
13& 
25& 
13& 
& 
14& 
45& 
48& 
& 
16& 
28& 
335 \\
\cline{1-3}
\cline{5-7}
\cline{9-11}
\cline{13-15}
9& 
19& 
6& 
& 
13& 
26& 
59& 
& 
14& 
46& 
9& 
& 
16& 
29& 
505 \\
\cline{1-3}
\cline{5-7}
\cline{9-11}
\cline{13-15}
9& 
21& 
1& 
& 
13& 
27& 
148& 
& 
14& 
47& 
1& 
& 
16& 
30& 
189 \\
\cline{1-3}
\cline{5-7}
\cline{9-11}
\cline{13-15}
9& 
24& 
3& 
& 
13& 
28& 
124& 
& 
14& 
48& 
16& 
& 
16& 
31& 
15 \\
\cline{1-3}
\cline{5-7}
\cline{9-11}
\cline{13-15}
10& 
16& 
3& 
& 
13& 
29& 
29& 
& 
14& 
49& 
6& 
& 
16& 
32& 
141 \\
\cline{1-3}
\cline{5-7}
\cline{9-11}
\cline{13-15}
10& 
18& 
1& 
& 
13& 
30& 
43& 
& 
15& 
19& 
24& 
& 
16& 
33& 
728 \\
\cline{1-3}
\cline{5-7}
\cline{9-11}
\cline{13-15}
10& 
22& 
9& 
& 
13& 
31& 
74& 
& 
15& 
20& 
29& 
& 
16& 
34& 
836 \\
\cline{1-3}
\cline{5-7}
\cline{9-11}
\cline{13-15}
10& 
24& 
9& 
& 
13& 
32& 
82& 
& 
15& 
21& 
26& 
& 
16& 
35& 
167 \\
\cline{1-3}
\cline{5-7}
\cline{9-11}
\cline{13-15}
10& 
26& 
7& 
& 
13& 
33& 
11& 
& 
15& 
22& 
102& 
& 
16& 
36& 
9 \\
\cline{1-3}
\cline{5-7}
\cline{9-11}
\cline{13-15}
11& 
11& 
2& 
& 
13& 
34& 
7& 
& 
15& 
23& 
1126& 
& 
16& 
37& 
16 \\
\cline{1-3}
\cline{5-7}
\cline{9-11}
\cline{13-15}
11& 
17& 
6& 
& 
13& 
35& 
19& 
& 
15& 
24& 
1392& 
& 
16& 
38& 
128 \\
\cline{1-3}
\cline{5-7}
\cline{9-11}
\cline{13-15}
11& 
19& 
1& 
& 
13& 
36& 
50& 
& 
15& 
25& 
473& 
& 
16& 
39& 
97 \\
\cline{1-3}
\cline{5-7}
\cline{9-11}
\cline{13-15}
11& 
21& 
12& 
& 
13& 
37& 
28& 
& 
15& 
26& 
76& 
& 
16& 
40& 
17 \\
\cline{1-3}
\cline{5-7}
\cline{9-11}
\cline{13-15}
11& 
22& 
2& 
& 
13& 
38& 
19& 
& 
15& 
27& 
1364& 
& 
16& 
41& 
34 \\
\cline{1-3}
\cline{5-7}
\cline{9-11}
\cline{13-15}

11& 
23& 
14& 
& 
13& 
39& 
20& 
& 
15& 
28& 
3963& 
& 
16& 
42& 
17 \\
\cline{1-3}
\cline{5-7}
\cline{9-11}
\cline{13-15}

11& 
24& 
2& 
& 
13& 
40& 
2& 
& 
15& 
29& 
2517& 
& 
16& 
43& 
16 \\
\cline{1-3}
\cline{5-7}
\cline{9-11}
\cline{13-15}

11& 
26& 
6& 
& 
13& 
41& 
4& 
& 
15& 
30& 
376& 
& 
16& 
44& 
31 \\
\cline{1-3}
\cline{5-7}
\cline{9-11}
\cline{13-15}

11& 
27& 
18& 
& 
13& 
43& 
2& 
& 
15& 
31& 
277& 
& 
16& 
45& 
22 \\
\cline{1-3}
\cline{5-7}
\cline{9-11}
\cline{13-15}

12& 
14& 
6& 
& 
14& 
16& 
7& 
& 
15& 
32& 
1653& 
& 
16& 
46& 
14 \\
\cline{1-3}
\cline{5-7}
\cline{9-11}
\cline{13-15}

12& 
16& 
8& 
& 
14& 
18& 
110& 
& 
15& 
33& 
2718& 
& 
16& 
47& 
17 \\
\cline{1-3}
\cline{5-7}
\cline{9-11}
\cline{13-15}

12& 
17& 
2& 
& 
14& 
19& 
132& 
& 
15& 
34& 
1047& 
& 
16& 
48& 
6 \\
\cline{1-3}
\cline{5-7}
\cline{9-11}
\cline{13-15}

12& 
18& 
21& 
& 
14& 
20& 
157& 
& 
15& 
35& 
60& 
& 
16& 
49& 
21 \\
\cline{1-3}
\cline{5-7}
\cline{9-11}
\cline{13-15}

12& 
21& 
14& 
& 
14& 
21& 
109& 
& 
15& 
36& 
115& 
& 
16& 
50& 
2 \\
\cline{1-3}
\cline{5-7}
\cline{9-11}
\cline{13-15}

12& 
22& 
27& 
& 
14& 
22& 
662& 
& 
15& 
37& 
317& 
& 
16& 
51& 
1 \\
\cline{1-3}
\cline{5-7}
\cline{9-11}
\cline{13-15}

12& 
23& 
1& 
& 
14& 
23& 
1735& 
& 
15& 
38& 
361& 
& 
16& 
53& 
2 \\
\cline{1-3}
\cline{5-7}
\cline{9-11}
\cline{13-15}

12& 
24& 
7& 
& 
14& 
24& 
1275& 
& 
15& 
39& 
137& 
& 
17& 
25& 
12 \\
\cline{1-3}
\cline{5-7}
\cline{9-11}
\cline{13-15}

12& 
26& 
17& 
& 
14& 
25& 
254& 
& 
15& 
40& 
174& 
& 
17& 
28& 
18 \\
\cline{1-3}
\cline{5-7}
\cline{9-11}
\cline{13-15}

12& 
27& 
8& 
& 
14& 
26& 
404& 
& 
15& 
41& 
222& 
& 
17& 
29& 
3 \\
\cline{1-3}
\cline{5-7}
\cline{9-11}
\cline{13-15}

12& 
28& 
4& 
& 
14& 
27& 
2081& 
& 
15& 
42& 
219& 
& 
17& 
32& 
9 \\
\cline{1-3}
\cline{5-7}
\cline{9-11}
\cline{13-15}

12& 
29& 
8& 
& 
14& 
28& 
3115& 
& 
15& 
43& 
195& 
& 
17& 
33& 
12 \\
\cline{1-3}
\cline{5-7}
\cline{9-11}
\cline{13-15}

12& 
30& 
2& 
& 
14& 
29& 
976& 
& 
15& 
44& 
238& 
& 
17& 
34& 
63 \\
\cline{1-3}
\cline{5-7}
\cline{9-11}
\cline{13-15}

12& 
31& 
6& 
& 
14& 
30& 
129& 
& 
15& 
45& 
98& 
& 
17& 
35& 
66 \\
\cline{1-3}
\cline{5-7}
\cline{9-11}
\cline{13-15}

12& 
32& 
10& 
& 
14& 
31& 
493& 
& 
15& 
46& 
63& 
& 
17& 
36& 
6 \\
\cline{1-3}
\cline{5-7}
\cline{9-11}
\cline{13-15}

12& 
33& 
4& 
& 
14& 
32& 
1389& 
& 
15& 
47& 
32& 
& 
17& 
39& 
45 \\
\cline{1-3}
\cline{5-7}
\cline{9-11}
\cline{13-15}

12& 
34& 
6& 
& 
14& 
33& 
702& 
& 
15& 
48& 
47& 
& 
17& 
40& 
9 \\
\cline{1-3}
\cline{5-7}
\cline{9-11}
\cline{13-15}

12& 
38& 
3& 
& 
14& 
34& 
46& 
& 
15& 
49& 
57& 
& 
18& 
29& 
3 \\
\cline{1-3}
\cline{5-7}
\cline{9-11}
\cline{13-15}

13& 
15& 
5& 
& 
14& 
35& 
86& 
& 
15& 
50& 
22& 
& 
18& 
33& 
6 \\
\cline{1-3}
\cline{5-7}
\cline{9-11}
\cline{13-15}

13& 
16& 
2& 
& 
14& 
36& 
137& 
& 
15& 
51& 
6& 
& 
18& 
36& 
6 \\
\cline{1-3}
\cline{5-7}
\cline{9-11}
\cline{13-15}

13& 
17& 
49& 
& 
14& 
37& 
263& 
& 
15& 
53& 
18& 
& 
18& 
40& 
9 \\
\cline{1-3}
\cline{5-7}
\cline{9-11}

13& 
18& 
52& 
& 
14& 
38& 
126& 
& 
16& 
20& 
3& 
& 
& 
& 
 \\
\cline{1-3}
\cline{5-7}
\cline{9-11}
\cline{13-15}

\end{tabular}
\caption{There are ${\#}specs$ specifications with maximum length= $MaxLen$ and cost = $cost$. }
\label{tab3}
\end{center}
\end{table}

Table \ref{tab2} shows the maximum length for the 40319 specifications, after excluding 
the identity. There are 3 specifications with maximum length = 7 gates and 24 specifications with maximum length = 18 gates. 
The average maximum length is 14.639. Table \ref{tab3} 
shows the same results as Table \ref{tab2} with more details on the cost of the maximum length, 
for example, there are 3 specifications with maximum length = 7 gates, the cost of two of them is 9,
 while the third specification has cost =11.

\begin{table}[htbp]
\small
\begin{center}
\begin{tabular}
{|p{48pt}|p{48pt}|p{75pt}|}
\hline
Min Len& 
{\#}specs& 
{\#}specs \cite{art} \\
\hline
1& 
12& 
12 \\
\hline
2& 
102& 
102 \\
\hline
3& 
625& 
625 \\
\hline
4& 
2780& 
2780 \\
\hline
5& 
8921& 
8921 \\
\hline
6& 
17049& 
17049 \\
\hline
7& 
10253& 
10253 \\
\hline
8& 
577& 
577 \\
\hline
Avg.& 
5.865& 
5.865 \\
\hline
\end{tabular}
\caption{There are ${\#}specs$ specifications with minimum length = $MinLen$.}
\label{tab4}
\end{center}
\end{table}

\begin{table}[htbp]
\small
\begin{center}
\begin{tabular}
{|p{15pt}|p{15pt}|p{25pt}|p{1pt}|p{15pt}|p{15pt}|p{25pt}|p{1pt}|p{15pt}|p{15pt}|p{25pt}|p{1pt}|p{15pt}|p{15pt}|p{25pt}|}
\cline{1-3}
\cline{5-7}
\cline{9-11}
\cline{13-15}
Min Len& 
cost& 
{\#}specs& 
& 
Min Len& 
cost& 
{\#}specs& 
& 
Min Len& 
cost& 
{\#}specs& 
& 
Min Len& 
cost& 
{\#}specs \\
\cline{1-3}
\cline{5-7}
\cline{9-11}
\cline{13-15}

1& 
0& 
3& 
& 
4& 
6& 
186& 
& 
5& 
18& 
5& 
& 
7& 
10& 
222 \\
\cline{1-3}
\cline{5-7}
\cline{9-11}
\cline{13-15}

1& 
1& 
6& 
& 
4& 
7& 
772& 
& 
5& 
19& 
45& 
& 
7& 
11& 
106 \\
\cline{1-3}
\cline{5-7}
\cline{9-11}
\cline{13-15}

1& 
5& 
3& 
& 
4& 
8& 
369& 
& 
5& 
20& 
14& 
& 
7& 
12& 
856 \\
\cline{1-3}
\cline{5-7}
\cline{9-11}
\cline{13-15}

2& 
0& 
3& 
& 
4& 
9& 
39& 
& 
5& 
21& 
1& 
& 
7& 
13& 
2278 \\
\cline{1-3}
\cline{5-7}
\cline{9-11}
\cline{13-15}

2& 
1& 
24& 
& 
4& 
10& 
228& 
& 
6& 
4& 
45& 
& 
7& 
14& 
1200 \\
\cline{1-3}
\cline{5-7}
\cline{9-11}
\cline{13-15}

2& 
2& 
24& 
& 
4& 
11& 
430& 
& 
6& 
5& 
168& 
& 
7& 
15& 
508 \\
\cline{1-3}
\cline{5-7}
\cline{9-11}
\cline{13-15}

2& 
5& 
15& 
& 
4& 
12& 
233& 
& 
6& 
6& 
2& 
& 
7& 
16& 
1493 \\
\cline{1-3}
\cline{5-7}
\cline{9-11}
\cline{13-15}

2& 
6& 
30& 
& 
4& 
14& 
45& 
& 
6& 
7& 
22& 
& 
7& 
17& 
1813 \\
\cline{1-3}
\cline{5-7}
\cline{9-11}
\cline{13-15}

2& 
9& 
3& 
& 
4& 
15& 
62& 
& 
6& 
8& 
1219& 
& 
7& 
18& 
848 \\
\cline{1-3}
\cline{5-7}
\cline{9-11}
\cline{13-15}

2& 
10& 
3& 
& 
4& 
16& 
13& 
& 
6& 
9& 
2233& 
& 
7& 
19& 
226 \\
\cline{1-3}
\cline{5-7}
\cline{9-11}
\cline{13-15}

3& 
0& 
1& 
& 
4& 
19& 
2& 
& 
6& 
10& 
253& 
& 
7& 
20& 
101 \\
\cline{1-3}
\cline{5-7}
\cline{9-11}
\cline{13-15}

3& 
1& 
18& 
& 
5& 
3& 
75& 
& 
6& 
11& 
833& 
& 
7& 
21& 
28 \\
\cline{1-3}
\cline{5-7}
\cline{9-11}
\cline{13-15}

3& 
2& 
117& 
& 
5& 
4& 
375& 
& 
6& 
12& 
3070& 
& 
7& 
22& 
3 \\
\cline{1-3}
\cline{5-7}
\cline{9-11}
\cline{13-15}

3& 
3& 
51& 
& 
5& 
5& 
24& 
& 
6& 
13& 
3933& 
& 
7& 
23& 
3 \\
\cline{1-3}
\cline{5-7}
\cline{9-11}
\cline{13-15}

3& 
5& 
24& 
& 
5& 
6& 
10& 
& 
6& 
14& 
759& 
& 
7& 
24& 
3 \\
\cline{1-3}
\cline{5-7}
\cline{9-11}
\cline{13-15}

3& 
6& 
162& 
& 
5& 
7& 
673& 
& 
6& 
15& 
913& 
& 
8& 
9& 
3 \\
\cline{1-3}
\cline{5-7}
\cline{9-11}
\cline{13-15}

3& 
7& 
138& 
& 
5& 
8& 
2041& 
& 
6& 
16& 
1860& 
& 
8& 
12& 
9 \\
\cline{1-3}
\cline{5-7}
\cline{9-11}
\cline{13-15}

3& 
9& 
18& 
& 
5& 
9& 
516& 
& 
6& 
17& 
1242& 
& 
8& 
13& 
31 \\
\cline{1-3}
\cline{5-7}
\cline{9-11}
\cline{13-15}

3& 
10& 
51& 
& 
5& 
10& 
284& 
& 
6& 
18& 
234& 
& 
8& 
14& 
36 \\
\cline{1-3}
\cline{5-7}
\cline{9-11}
\cline{13-15}

3& 
11& 
39& 
& 
5& 
11& 
1181& 
& 
6& 
19& 
142& 
& 
8& 
15& 
22 \\
\cline{1-3}
\cline{5-7}
\cline{9-11}
\cline{13-15}

3& 
14& 
5& 
& 
5& 
12& 
1883& 
& 
6& 
20& 
104& 
& 
8& 
16& 
145 \\
\cline{1-3}
\cline{5-7}
\cline{9-11}
\cline{13-15}

3& 
15& 
1& 
& 
5& 
13& 
669& 
& 
6& 
21& 
16& 
& 
8& 
17& 
173 \\
\cline{1-3}
\cline{5-7}
\cline{9-11}
\cline{13-15}

4& 
2& 
51& 
& 
5& 
14& 
137& 
& 
6& 
22& 
1& 
& 
8& 
18& 
109 \\
\cline{1-3}
\cline{5-7}
\cline{9-11}
\cline{13-15}

4& 
3& 
282& 
& 
5& 
15& 
490& 
& 
7& 
6& 
14& 
& 
8& 
19& 
39 \\
\cline{1-3}
\cline{5-7}
\cline{9-11}
\cline{13-15}

4& 
4& 
60& 
& 
5& 
16& 
409& 
& 
7& 
8& 
33& 
& 
8& 
20& 
9 \\
\cline{1-3}
\cline{5-7}
\cline{9-11}
\cline{13-15}

4& 
5& 
8& 
& 
5& 
17& 
89& 
& 
7& 
9& 
518& 
& 
8& 
21& 
1 \\
\cline{1-3}
\cline{5-7}
\cline{9-11}
\cline{13-15}

\end{tabular}
\caption{There are ${\#}specs$ specifications with minimum length= $MinLen$ and cost = $cost$.}
\label{tab5}
\end{center}
\end{table}

Table \ref{tab4} shows the minimum length for the 40319 specifications, after excluding 
the identity. There are 12 specifications with minimum length = 1 gate, 
these are the generators shown in Eqn.\ref{12gatecycle}, while there are 
577 specifications with minimum length = 8 gates. These results are identical with that 
shown in \cite{art}. This comes from the fact that all minimum length circuits come from 
using the gate library that contains all the 12 generators shown in Eqn.\ref{12gatecycle}. 
This is not true when finding a library of gates that synthesize minimum cost circuits as shown later.
Table \ref{tab5} shows the same results as Table \ref{tab4} with more details 
on the cost of the minimum length circuit.

\begin{table}[htbp]
\small
\begin{center}
\begin{tabular}
{|p{15pt}|p{25pt}|p{1pt}|p{15pt}|p{25pt}|p{1pt}|p{15pt}|p{25pt}|}
\cline{1-2}\cline{4-5}\cline{7-8}
Max Cost& 
{\#}specs& 
& 
Max Cost& 
{\#}specs& 
& 
Max Cost& 
{\#}specs \\
\cline{1-2}\cline{4-5}\cline{7-8}
18& 
1& 
& 
31& 
122& 
& 
43& 
5406 \\
\cline{1-2}\cline{4-5}\cline{7-8}
19& 
1& 
& 
32& 
314& 
& 
44& 
3484 \\
\cline{1-2}\cline{4-5}\cline{7-8}
20& 
1& 
& 
33& 
549& 
& 
45& 
1652 \\
\cline{1-2}\cline{4-5}\cline{7-8}
22& 
10& 
& 
34& 
525& 
& 
46& 
883 \\
\cline{1-2}\cline{4-5}\cline{7-8}
23& 
9& 
& 
35& 
972& 
& 
47& 
687 \\
\cline{1-2}\cline{4-5}\cline{7-8}
24& 
21& 
& 
36& 
1816& 
& 
48& 
914 \\
\cline{1-2}\cline{4-5}\cline{7-8}
25& 
22& 
& 
37& 
3088& 
& 
49& 
622 \\
\cline{1-2}\cline{4-5}\cline{7-8}
26& 
18& 
& 
38& 
3035& 
& 
50& 
96 \\
\cline{1-2}\cline{4-5}\cline{7-8}
27& 
41& 
& 
39& 
3769& 
& 
51& 
28 \\
\cline{1-2}\cline{4-5}\cline{7-8}
28& 
23& 
& 
40& 
4111& 
& 
52& 
5 \\
\cline{1-2}\cline{4-5}\cline{7-8}
29& 
37& 
& 
41& 
3854& 
& 
53& 
68 \\
\cline{1-2}\cline{4-5}\cline{7-8}
30& 
66& 
& 
42& 
4067& 
& 
54& 
2 \\
\cline{1-2}\cline{4-5}\cline{7-8}
\end{tabular}
\caption{There are ${\#}specs$ specifications with maximum cost = $MaxCost$. The average maximum cost is 40.759. }
\label{tab6}
\end{center}
\end{table}

\begin{table}[htbp]
\small
\begin{center}
\begin{tabular}
{|p{15pt}|p{15pt}|p{25pt}|p{1pt}|p{15pt}|p{15pt}|p{25pt}|p{1pt}|p{15pt}|p{15pt}|p{25pt}|p{1pt}|p{15pt}|p{15pt}|p{25pt}|}
\cline{1-3}
\cline{5-7}
\cline{9-11}
\cline{13-15}

Max Cost& 
Len& 
{\#}specs& 
& 
Max Cost& 
Len& 
{\#}specs& 
& 
Max Cost& 
Len& 
{\#}specs& 
& 
Max Cost& 
Len& 
{\#}specs \\
\cline{1-3}
\cline{5-7}
\cline{9-11}
\cline{13-15}

18& 
6& 
1& 
& 
30& 
13& 
1& 
& 
37& 
11& 
207& 
& 
44& 
14& 
1652 \\
\cline{1-3}
\cline{5-7}
\cline{9-11}
\cline{13-15}

19& 
6& 
1& 
& 
31& 
7& 
1& 
& 
37& 
12& 
575& 
& 
44& 
13& 
946 \\
\cline{1-3}
\cline{5-7}
\cline{9-11}
\cline{13-15}

20& 
6& 
1& 
& 
31& 
9& 
1& 
& 
37& 
13& 
1268& 
& 
44& 
15& 
494 \\
\cline{1-3}
\cline{5-7}
\cline{9-11}
\cline{13-15}

22& 
7& 
1& 
& 
31& 
10& 
25& 
& 
37& 
14& 
864& 
& 
44& 
16& 
28 \\
\cline{1-3}
\cline{5-7}
\cline{9-11}
\cline{13-15}

22& 
10& 
3& 
& 
31& 
11& 
29& 
& 
37& 
15& 
127& 
& 
44& 
12& 
351 \\
\cline{1-3}
\cline{5-7}
\cline{9-11}
\cline{13-15}

22& 
11& 
3& 
& 
31& 
12& 
33& 
& 
37& 
16& 
3& 
& 
44& 
11& 
13 \\
\cline{1-3}
\cline{5-7}
\cline{9-11}
\cline{13-15}

22& 
12& 
3& 
& 
31& 
13& 
27& 
& 
38& 
10& 
72& 
& 
45& 
11& 
1 \\
\cline{1-3}
\cline{5-7}
\cline{9-11}
\cline{13-15}

23& 
7& 
2& 
& 
31& 
14& 
5& 
& 
38& 
11& 
376& 
& 
45& 
12& 
98 \\
\cline{1-3}
\cline{5-7}
\cline{9-11}
\cline{13-15}

23& 
8& 
4& 
& 
31& 
15& 
1& 
& 
38& 
12& 
745& 
& 
45& 
13& 
215 \\
\cline{1-3}
\cline{5-7}
\cline{9-11}
\cline{13-15}

23& 
11& 
2& 
& 
32& 
8& 
1& 
& 
38& 
13& 
972& 
& 
45& 
14& 
921 \\
\cline{1-3}
\cline{5-7}
\cline{9-11}
\cline{13-15}

23& 
12& 
1& 
& 
32& 
10& 
36& 
& 
38& 
14& 
642& 
& 
45& 
15& 
394 \\
\cline{1-3}
\cline{5-7}
\cline{9-11}
\cline{13-15}

24& 
7& 
4& 
& 
32& 
11& 
53& 
& 
38& 
15& 
181& 
& 
45& 
16& 
23 \\
\cline{1-3}
\cline{5-7}
\cline{9-11}
\cline{13-15}

24& 
8& 
7& 
& 
32& 
12& 
98& 
& 
38& 
16& 
47& 
& 
46& 
12& 
31 \\
\cline{1-3}
\cline{5-7}
\cline{9-11}
\cline{13-15}

24& 
9& 
4& 
& 
32& 
13& 
73& 
& 
39& 
10& 
24& 
& 
46& 
13& 
109 \\
\cline{1-3}
\cline{5-7}
\cline{9-11}
\cline{13-15}

24& 
10& 
3& 
& 
32& 
14& 
44& 
& 
39& 
11& 
305& 
& 
46& 
14& 
405 \\
\cline{1-3}
\cline{5-7}
\cline{9-11}
\cline{13-15}

24& 
11& 
2& 
& 
32& 
15& 
9& 
& 
39& 
12& 
983& 
& 
46& 
15& 
323 \\
\cline{1-3}
\cline{5-7}
\cline{9-11}
\cline{13-15}

24& 
12& 
1& 
& 
33& 
8& 
2& 
& 
39& 
13& 
1467& 
& 
46& 
16& 
15 \\
\cline{1-3}
\cline{5-7}
\cline{9-11}
\cline{13-15}

25& 
7& 
18& 
& 
33& 
9& 
15& 
& 
39& 
14& 
830& 
& 
47& 
12& 
22 \\
\cline{1-3}
\cline{5-7}
\cline{9-11}
\cline{13-15}

25& 
8& 
2& 
& 
33& 
10& 
18& 
& 
39& 
15& 
132& 
& 
47& 
13& 
168 \\
\cline{1-3}
\cline{5-7}
\cline{9-11}
\cline{13-15}

25& 
9& 
2& 
& 
33& 
11& 
54& 
& 
39& 
16& 
28& 
& 
47& 
14& 
275 \\
\cline{1-3}
\cline{5-7}
\cline{9-11}
\cline{13-15}

26& 
7& 
5& 
& 
33& 
12& 
193& 
& 
40& 
10& 
2& 
& 
47& 
15& 
204 \\
\cline{1-3}
\cline{5-7}
\cline{9-11}
\cline{13-15}

26& 
8& 
2& 
& 
33& 
13& 
173& 
& 
40& 
11& 
83& 
& 
47& 
16& 
18 \\
\cline{1-3}
\cline{5-7}
\cline{9-11}
\cline{13-15}

26& 
9& 
1& 
& 
33& 
14& 
51& 
& 
40& 
12& 
631& 
& 
48& 
13& 
147 \\
\cline{1-3}
\cline{5-7}
\cline{9-11}
\cline{13-15}

26& 
10& 
10& 
& 
33& 
15& 
43& 
& 
40& 
13& 
1682& 
& 
48& 
14& 
529 \\
\cline{1-3}
\cline{5-7}
\cline{9-11}
\cline{13-15}

27& 
7& 
14& 
& 
34& 
8& 
1& 
& 
40& 
14& 
1437& 
& 
48& 
15& 
230 \\
\cline{1-3}
\cline{5-7}
\cline{9-11}
\cline{13-15}

27& 
8& 
2& 
& 
34& 
9& 
1& 
& 
40& 
15& 
261& 
& 
48& 
16& 
8 \\
\cline{1-3}
\cline{5-7}
\cline{9-11}
\cline{13-15}

27& 
9& 
5& 
& 
34& 
10& 
16& 
& 
40& 
16& 
15& 
& 
49& 
13& 
1 \\
\cline{1-3}
\cline{5-7}
\cline{9-11}
\cline{13-15}

27& 
11& 
18& 
& 
34& 
11& 
88& 
& 
41& 
11& 
58& 
& 
49& 
14& 
296 \\
\cline{1-3}
\cline{5-7}
\cline{9-11}
\cline{13-15}

27& 
12& 
2& 
& 
34& 
12& 
186& 
& 
41& 
12& 
378& 
& 
49& 
15& 
303 \\
\cline{1-3}
\cline{5-7}
\cline{9-11}
\cline{13-15}

28& 
8& 
8& 
& 
34& 
13& 
168& 
& 
41& 
13& 
1297& 
& 
49& 
16& 
22 \\
\cline{1-3}
\cline{5-7}
\cline{9-11}
\cline{13-15}

28& 
9& 
3& 
& 
34& 
14& 
41& 
& 
41& 
14& 
1685& 
& 
50& 
13& 
1 \\
\cline{1-3}
\cline{5-7}
\cline{9-11}
\cline{13-15}

28& 
10& 
3& 
& 
34& 
15& 
24& 
& 
41& 
15& 
411& 
& 
50& 
14& 
10 \\
\cline{1-3}
\cline{5-7}
\cline{9-11}
\cline{13-15}

28& 
11& 
6& 
& 
35& 
9& 
2& 
& 
41& 
16& 
25& 
& 
50& 
15& 
83 \\
\cline{1-3}
\cline{5-7}
\cline{9-11}
\cline{13-15}

28& 
12& 
3& 
& 
35& 
10& 
8& 
& 
42& 
11& 
107& 
& 
50& 
16& 
2 \\
\cline{1-3}
\cline{5-7}
\cline{9-11}
\cline{13-15}

29& 
7& 
2& 
& 
35& 
11& 
127& 
& 
42& 
12& 
854& 
& 
51& 
14& 
9 \\
\cline{1-3}
\cline{5-7}
\cline{9-11}
\cline{13-15}

29& 
8& 
1& 
& 
35& 
12& 
370& 
& 
42& 
13& 
1391& 
& 
51& 
15& 
18 \\
\cline{1-3}
\cline{5-7}
\cline{9-11}
\cline{13-15}

29& 
9& 
4& 
& 
35& 
13& 
406& 
& 
42& 
14& 
1305& 
& 
51& 
16& 
1 \\
\cline{1-3}
\cline{5-7}
\cline{9-11}
\cline{13-15}

29& 
10& 
7& 
& 
35& 
14& 
59& 
& 
42& 
15& 
393& 
& 
52& 
14& 
1 \\
\cline{1-3}
\cline{5-7}
\cline{9-11}
\cline{13-15}

29& 
11& 
6& 
& 
36& 
10& 
2& 
& 
42& 
16& 
17& 
& 
52& 
15& 
4 \\
\cline{1-3}
\cline{5-7}
\cline{9-11}
\cline{13-15}

29& 
12& 
17& 
& 
36& 
11& 
98& 
& 
43& 
11& 
87& 
& 
53& 
14& 
22 \\
\cline{1-3}
\cline{5-7}
\cline{9-11}
\cline{13-15}

30& 
8& 
2& 
& 
36& 
12& 
556& 
& 
43& 
12& 
1251& 
& 
53& 
15& 
44 \\
\cline{1-3}
\cline{5-7}
\cline{9-11}
\cline{13-15}

30& 
9& 
3& 
& 
36& 
13& 
839& 
& 
43& 
13& 
2241& 
& 
53& 
16& 
2 \\
\cline{1-3}
\cline{5-7}
\cline{9-11}
\cline{13-15}

30& 
10& 
38& 
& 
36& 
14& 
292& 
& 
43& 
14& 
1457& 
& 
54& 
15& 
2 \\
\cline{1-3}
\cline{5-7}
\cline{9-11}

30& 
11& 
7& 
& 
36& 
15& 
29& 
& 
43& 
15& 
355& 
& 
& 
& 
 \\
\cline{1-3}
\cline{5-7}
\cline{9-11}

30& 
12& 
15& 
& 
37& 
10& 
44& 
& 
43& 
16& 
15& 
& 
& 
& 
 \\
\cline{1-3}
\cline{5-7}
\cline{9-11}
\cline{13-15}

\end{tabular}
\caption{There are ${\#}specs$ specifications with maximum cost = $MaxCost$ and 
length = $Len$.}
\label{tab7}
\end{center}
\end{table}

Table \ref{tab6} shows the maximum cost for the 40319 specifications, after excluding 
the identity. There is 1 specification with maximum cost = 18 and 
2 specifications with maximum cost = 54. 
The average maximum cost is 40.759. Table \ref{tab7} 
shows the same results as Table \ref{tab2} with more details on the length of the maximum cost, 
for example, the length of the only specification with maximum cost = 18 is 6 gates and the length 
of the two specifications with maximum cost = 54 are 15 gates.

\begin{table}[htbp]
\small
\begin{center}
\begin{tabular}
{|p{25pt}|p{35pt}|p{40pt}|}
\hline
Min Cost& 
{\#}specs& 
{\#}specs\cite{art} \\
\hline
0& 
7& 
7 \\
\hline
1& 
48& 
48 \\
\hline
2& 
192& 
192 \\
\hline
3& 
408& 
408 \\
\hline
4& 
480& 
480 \\
\hline
5& 
288& 
288 \\
\hline
6& 
592& 
592 \\
\hline
7& 
1962& 
2016 \\
\hline
8& 
3887& 
4128 \\
\hline
9& 
2916& 
2496 \\
\hline
10& 
1299& 
672 \\
\hline
11& 
3683& 
2880 \\
\hline
12& 
7221& 
7488 \\
\hline
13& 
6059& 
7488 \\
\hline
14& 
1465& 
384 \\
\hline
15& 
3562& 
1600 \\
\hline
16& 
4201& 
5568 \\
\hline
17& 
2049& 
3584 \\
\hline
Avg.& 
11.76967& 
11.98313 \\
\hline
\end{tabular}
\caption{There are ${\#}specs$ specifications with minimum cost = $MinCost$.}
\label{tab8}
\end{center}
\end{table}

Table \ref{tab8} shows the minimum cost for the 40319 specifications, after excluding 
the identity. There are 7 specifications with minimum cost = 0, these are the cost 
of specifications synthesized by circuits contain the gates $N_1$, $N_2$ and $N_3$ only. 
There are 7 such circuits starting from a circuit with only one $N$ gate to a circuit consists 
of the three gates: $(N_1 N_2 N_3)$.
There are 2049 specifications with minimum cost = 17. 
The average minimum cost is 11.769. This is better that that shown in \cite{art}, 
this enhancement comes from considering all sub-libraries. Table \ref{tab9} 
shows the same results as Table \ref{tab8} with more details on the length of the circuits with minimum cost.

\begin{table}[htbp]
\small
\begin{center}
\begin{tabular}
{|p{15pt}|p{15pt}|p{25pt}|p{1pt}|p{15pt}|p{15pt}|p{25pt}|p{1pt}|p{15pt}|p{15pt}|p{25pt}|p{1pt}|p{15pt}|p{15pt}|p{25pt}|}
\cline{1-3}
\cline{5-7}
\cline{9-11}
\cline{13-15}
Min Cost& 
Len& 
{\#}specs& 
& 
Min Cost& 
Len& 
{\#}specs& 
& 
Min Cost& 
Len& 
{\#}specs& 
& 
Min Cost& 
Len& 
{\#}specs \\
\cline{1-3}
\cline{5-7}
\cline{9-11}
\cline{13-15}
0& 
1& 
3& 
& 
7& 
4& 
615& 
& 
11& 
4& 
261& 
& 
14& 
11& 
13 \\
\cline{1-3}
\cline{5-7}
\cline{9-11}
\cline{13-15}
0& 
2& 
3& 
& 
7& 
5& 
820& 
& 
11& 
5& 
962& 
& 
14& 
12& 
2 \\
\cline{1-3}
\cline{5-7}
\cline{9-11}
\cline{13-15}
0& 
3& 
1& 
& 
7& 
6& 
362& 
& 
11& 
6& 
1366& 
& 
15& 
3& 
1 \\
\cline{1-3}
\cline{5-7}
\cline{9-11}
\cline{13-15}
1& 
1& 
6& 
& 
7& 
7& 
63& 
& 
11& 
7& 
801& 
& 
15& 
4& 
35 \\
\cline{1-3}
\cline{5-7}
\cline{9-11}
\cline{13-15}
1& 
2& 
24& 
& 
7& 
8& 
3& 
& 
11& 
8& 
248& 
& 
15& 
5& 
295 \\
\cline{1-3}
\cline{5-7}
\cline{9-11}
\cline{13-15}
1& 
3& 
18& 
& 
8& 
4& 
249& 
& 
11& 
9& 
18& 
& 
15& 
6& 
869 \\
\cline{1-3}
\cline{5-7}
\cline{9-11}
\cline{13-15}
2& 
2& 
24& 
& 
8& 
5& 
1649& 
& 
12& 
4& 
123& 
& 
15& 
7& 
1111 \\
\cline{1-3}
\cline{5-7}
\cline{9-11}
\cline{13-15}
2& 
3& 
117& 
& 
8& 
6& 
1489& 
& 
12& 
5& 
1090& 
& 
15& 
8& 
805 \\
\cline{1-3}
\cline{5-7}
\cline{9-11}
\cline{13-15}
2& 
4& 
51& 
& 
8& 
7& 
446& 
& 
12& 
6& 
2738& 
& 
15& 
9& 
383 \\
\cline{1-3}
\cline{5-7}
\cline{9-11}
\cline{13-15}
3& 
3& 
51& 
& 
8& 
8& 
48& 
& 
12& 
7& 
2233& 
& 
15& 
10& 
63 \\
\cline{1-3}
\cline{5-7}
\cline{9-11}
\cline{13-15}
3& 
4& 
282& 
& 
8& 
9& 
6& 
& 
12& 
8& 
869& 
& 
16& 
4& 
8 \\
\cline{1-3}
\cline{5-7}
\cline{9-11}
\cline{13-15}
3& 
5& 
75& 
& 
9& 
2& 
3& 
& 
12& 
9& 
165& 
& 
16& 
5& 
154 \\
\cline{1-3}
\cline{5-7}
\cline{9-11}
\cline{13-15}
4& 
4& 
60& 
& 
9& 
3& 
18& 
& 
12& 
10& 
3& 
& 
16& 
6& 
873 \\
\cline{1-3}
\cline{5-7}
\cline{9-11}
\cline{13-15}
4& 
5& 
375& 
& 
9& 
4& 
45& 
& 
13& 
5& 
276& 
& 
16& 
7& 
1443 \\
\cline{1-3}
\cline{5-7}
\cline{9-11}
\cline{13-15}
4& 
6& 
45& 
& 
9& 
5& 
356& 
& 
13& 
6& 
2244& 
& 
16& 
8& 
1023 \\
\cline{1-3}
\cline{5-7}
\cline{9-11}
\cline{13-15}
5& 
1& 
3& 
& 
9& 
6& 
1794& 
& 
13& 
7& 
2158& 
& 
16& 
9& 
591 \\
\cline{1-3}
\cline{5-7}
\cline{9-11}
\cline{13-15}
5& 
2& 
15& 
& 
9& 
7& 
600& 
& 
13& 
8& 
875& 
& 
16& 
10& 
109 \\
\cline{1-3}
\cline{5-7}
\cline{9-11}
\cline{13-15}
5& 
3& 
30& 
& 
9& 
8& 
97& 
& 
13& 
9& 
381& 
& 
17& 
5& 
26 \\
\cline{1-3}
\cline{5-7}
\cline{9-11}
\cline{13-15}
5& 
4& 
30& 
& 
9& 
9& 
2& 
& 
13& 
10& 
113& 
& 
17& 
6& 
290 \\
\cline{1-3}
\cline{5-7}
\cline{9-11}
\cline{13-15}
5& 
5& 
39& 
& 
9& 
10& 
1& 
& 
13& 
11& 
12& 
& 
17& 
7& 
548 \\
\cline{1-3}
\cline{5-7}
\cline{9-11}
\cline{13-15}
5& 
6& 
171& 
& 
10& 
2& 
3& 
& 
14& 
3& 
5& 
& 
17& 
8& 
332 \\
\cline{1-3}
\cline{5-7}
\cline{9-11}
\cline{13-15}
6& 
2& 
24& 
& 
10& 
3& 
39& 
& 
14& 
4& 
41& 
& 
17& 
9& 
347 \\
\cline{1-3}
\cline{5-7}
\cline{9-11}
\cline{13-15}
6& 
3& 
132& 
& 
10& 
4& 
180& 
& 
14& 
5& 
141& 
& 
17& 
10& 
338 \\
\cline{1-3}
\cline{5-7}
\cline{9-11}
\cline{13-15}
6& 
4& 
222& 
& 
10& 
5& 
363& 
& 
14& 
6& 
360& 
& 
17& 
11& 
155 \\
\cline{1-3}
\cline{5-7}
\cline{9-11}
\cline{13-15}
6& 
5& 
150& 
& 
10& 
6& 
391& 
& 
14& 
7& 
473& 
& 
17& 
12& 
13 \\
\cline{1-3}
\cline{5-7}
\cline{9-11}
6& 
6& 
50& 
& 
10& 
7& 
295& 
& 
14& 
8& 
247& 
& 
& 
& 
 \\
\cline{1-3}
\cline{5-7}
\cline{9-11}
6& 
7& 
14& 
& 
10& 
8& 
28& 
& 
14& 
9& 
125& 
& 
& 
& 
 \\
\cline{1-3}
\cline{5-7}
\cline{9-11}
7& 
3& 
99& 
& 
11& 
3& 
27& 
& 
14& 
10& 
58& 
& 
& 
& 
 \\
\cline{1-3}
\cline{5-7}
\cline{9-11}
\cline{13-15}
\end{tabular}
\caption{There are ${\#}specs$ specifications with minimum cost = $MinCost$ and 
length = $Len$.}
\label{tab9}
\end{center}
\end{table}

Comparing the above results related to the bounds on circuits length with 
the results related to the bound on circuits cost shows that 
the cost and the length of a circuit using a library is not correlated if all 
types of generators are considered, not only the gates used to calculate the cost, 
e.g. $T$ gate is counted as 1 in circuit length while it is counted 5 in circuit cost. 
For example, when searching for the best length circuit for 
the specification $((1,7)(2,5)(3,6,8,4))$, the best length circuit is 
$( F_{12}  T_{321}  T_{123}  N_1  F_{13}  F_{32}  T_{123} )$ with length = 7 and quantum cost = 18, while 
searching for the best cost circuit for the same specification, 
the best cost circuit is $( N_2  N_3  T_{321}  N_2  F_{13}  T_{321}  N_2  N_3  F_{12}  T_{321}  N_2  N_3 )$ 
with length = 12 and cost = 17.

\subsection{Library Oriented Results}
\label{lor}

In this section, the results are organized to show a summary of library oriented results 
in Table \ref{tab10}. 
The results related to the bounds on circuits maximum length synthesized by the sub-libraries
are shown in Tables \ref{tab11}, \ref{tab12} and \ref{tab13}.
Then results related to the bounds on circuits maximum cost synthesized by the sub-libraries 
are shown in Tables \ref{tab14}, \ref{tab15} and \ref{tab16}. The results of minimum cost and minimum circuit length are trivial results, since all sub-libraries 
can synthesize a circuit of length 1 and cost 0, 1 or 5.

\begin{table}[htbp]
\small
\begin{center}
\begin{tabular}
{|p{48pt}|p{48pt}|p{1pt}|p{48pt}|p{48pt}|} \cline{1-2}\cline{4-5}
{\#}lib & {\#}specs& &{\#}lib & {\#}specs \\\cline{1-2}\cline{4-5}
1& 4 &&	90& 15	\\\cline{1-2}\cline{4-5}
1& 1342&& 90& 31	\\\cline{1-2}\cline{4-5}
5& 5 &&	90& 47	\\\cline{1-2}\cline{4-5}
9& 35&& 96& 63	\\\cline{1-2}\cline{4-5}
12&1&& 	96& 127	\\\cline{1-2}\cline{4-5}
15&11&& 99& 95	\\\cline{1-2}\cline{4-5}
18&167&& 108& 575	\\\cline{1-2}\cline{4-5}
24&71&& 117& 719	\\\cline{1-2}\cline{4-5}
24&119&& 125& 1343	\\\cline{1-2}\cline{4-5}
27&143&& 162& 5039	\\\cline{1-2}\cline{4-5}
30&3&& 	168& 1439	\\\cline{1-2}\cline{4-5}
37&23&& 186& 191	\\\cline{1-2}\cline{4-5}
72&383&& 360& 1151	\\\cline{1-2}\cline{4-5}
73& 7&& 1960& 40319	\\\cline{1-2}\cline{4-5}
\end{tabular}
\caption{There are ${\#}Lib$ sub-libraries that can synthesize a circuit for ${\#}specs$ 
specifications.}
\label{tab10}
\end{center}
\end{table}

Table \ref{tab10} shows the ability of a sub-library to synthesize 
a circuit for a specification. It can be seen that there are 1960 sub-libraries 
that can be used to synthesize a circuit of any specification, i.e. there are 1960 universal reversible 
sub-libraries from the main library shown in Eqn.\ref{12gatecycle}. It is shown 
in the previous section that the main library is the 
best library to synthesize a minimum length circuit, while choosing a 
sub-library to synthesize a minimal cost circuit is not a trivial task.

\begin{table}[htbp]
\small
\begin{center}
\begin{tabular}
{|p{32pt}|p{29pt}|p{1pt}|p{29pt}|p{29pt}|}\cline{1-2}\cline{4-5}
Max Len & {\#}lib&&Max Len & {\#}lib \\\cline{1-2}\cline{4-5}
1& 12  &&10& 498 \\\cline{1-2}\cline{4-5}
2& 54  &&11& 498 \\\cline{1-2}\cline{4-5}
3& 79  &&12& 433 \\\cline{1-2}\cline{4-5}
4& 148 &&13& 378 \\\cline{1-2}\cline{4-5}
5& 231 &&14& 255 \\\cline{1-2}\cline{4-5}
6& 289 &&15& 120 \\\cline{1-2}\cline{4-5}
7& 223 &&16& 51 \\\cline{1-2}\cline{4-5}
8& 364 &&17& 6 \\\cline{1-2}\cline{4-5}
9& 453 &&18& 3 \\\cline{1-2}\cline{4-5}

\end{tabular}
\caption{There are ${\#}lib$ sub-libraries that can synthesize 
circuits with maximum length = $MaxLen$.}
\label{tab11}
\end{center}
\end{table}

\begin{table}[htbp]
\small
\begin{center}
\begin{tabular}
{|p{15pt}|p{15pt}|p{25pt}|p{1pt}|p{15pt}|p{15pt}|p{25pt}|p{1pt}|p{15pt}|p{15pt}|p{25pt}|p{1pt}|p{15pt}|p{15pt}|p{25pt}|}
\cline{1-3}
\cline{5-7}
\cline{9-11}
\cline{13-15}
Max Len& 
cost& 
{\#}lib& 
& 
Max Len& 
cost& 
{\#}lib& 
& 
Max Len& 
cost& 
{\#}lib& 
& 
Max Len& 
cost& 
{\#}lib \\
\cline{1-3}
\cline{5-7}
\cline{9-11}
\cline{13-15}
1& 
0& 
3& 
& 
6& 
7& 
15& 
& 
8& 
19& 
42& 
& 
10& 
21& 
14 \\
\cline{1-3}
\cline{5-7}
\cline{9-11}
\cline{13-15}
1& 
1& 
6& 
& 
6& 
8& 
26& 
& 
8& 
20& 
22& 
& 
10& 
22& 
30 \\
\cline{1-3}
\cline{5-7}
\cline{9-11}
\cline{13-15}
1& 
5& 
3& 
& 
6& 
9& 
21& 
& 
8& 
21& 
1& 
& 
10& 
23& 
26 \\
\cline{1-3}
\cline{5-7}
\cline{9-11}
\cline{13-15}
2& 
0& 
3& 
& 
6& 
10& 
36& 
& 
8& 
22& 
10& 
& 
10& 
24& 
35 \\
\cline{1-3}
\cline{5-7}
\cline{9-11}
\cline{13-15}
2& 
1& 
18& 
& 
6& 
11& 
23& 
& 
8& 
23& 
10& 
& 
10& 
25& 
4 \\
\cline{1-3}
\cline{5-7}
\cline{9-11}
\cline{13-15}
2& 
2& 
12& 
& 
6& 
12& 
38& 
& 
8& 
24& 
13& 
& 
10& 
26& 
17 \\
\cline{1-3}
\cline{5-7}
\cline{9-11}
\cline{13-15}
2& 
5& 
3& 
& 
6& 
13& 
5& 
& 
8& 
25& 
3& 
& 
10& 
27& 
3 \\
\cline{1-3}
\cline{5-7}
\cline{9-11}
\cline{13-15}
2& 
6& 
18& 
& 
6& 
14& 
11& 
& 
8& 
27& 
1& 
& 
10& 
28& 
21 \\
\cline{1-3}
\cline{5-7}
\cline{9-11}
\cline{13-15}
3& 
0& 
1& 
& 
6& 
15& 
3& 
& 
8& 
28& 
2& 
& 
10& 
29& 
9 \\
\cline{1-3}
\cline{5-7}
\cline{9-11}
\cline{13-15}
3& 
1& 
12& 
& 
6& 
16& 
3& 
& 
8& 
30& 
1& 
& 
10& 
30& 
19 \\
\cline{1-3}
\cline{5-7}
\cline{9-11}
\cline{13-15}
3& 
2& 
21& 
& 
6& 
18& 
8& 
& 
9& 
7& 
48& 
& 
10& 
32& 
2 \\
\cline{1-3}
\cline{5-7}
\cline{9-11}
\cline{13-15}
3& 
3& 
3& 
& 
6& 
20& 
2& 
& 
9& 
8& 
20& 
& 
10& 
34& 
5 \\
\cline{1-3}
\cline{5-7}
\cline{9-11}
\cline{13-15}
3& 
6& 
24& 
& 
6& 
21& 
2& 
& 
9& 
11& 
10& 
& 
10& 
35& 
1 \\
\cline{1-3}
\cline{5-7}
\cline{9-11}
\cline{13-15}
3& 
7& 
15& 
& 
6& 
22& 
2& 
& 
9& 
12& 
3& 
& 
10& 
36& 
2 \\
\cline{1-3}
\cline{5-7}
\cline{9-11}
\cline{13-15}
3& 
14& 
2& 
& 
7& 
5& 
39& 
& 
9& 
13& 
19& 
& 
11& 
13& 
4 \\
\cline{1-3}
\cline{5-7}
\cline{9-11}
\cline{13-15}
3& 
15& 
1& 
& 
7& 
6& 
10& 
& 
9& 
14& 
44& 
& 
11& 
14& 
1 \\
\cline{1-3}
\cline{5-7}
\cline{9-11}
\cline{13-15}
4& 
2& 
40& 
& 
7& 
7& 
6& 
& 
9& 
15& 
31& 
& 
11& 
15& 
1 \\
\cline{1-3}
\cline{5-7}
\cline{9-11}
\cline{13-15}
4& 
3& 
9& 
& 
7& 
8& 
14& 
& 
9& 
16& 
6& 
& 
11& 
16& 
45 \\
\cline{1-3}
\cline{5-7}
\cline{9-11}
\cline{13-15}
4& 
4& 
33& 
& 
7& 
9& 
27& 
& 
9& 
17& 
40& 
& 
11& 
17& 
54 \\
\cline{1-3}
\cline{5-7}
\cline{9-11}
\cline{13-15}
4& 
6& 
12& 
& 
7& 
11& 
12& 
& 
9& 
18& 
49& 
& 
11& 
18& 
20 \\
\cline{1-3}
\cline{5-7}
\cline{9-11}
\cline{13-15}
4& 
7& 
17& 
& 
7& 
12& 
4& 
& 
9& 
19& 
37& 
& 
11& 
19& 
47 \\
\cline{1-3}
\cline{5-7}
\cline{9-11}
\cline{13-15}
4& 
8& 
7& 
& 
7& 
13& 
22& 
& 
9& 
20& 
20& 
& 
11& 
20& 
42 \\
\cline{1-3}
\cline{5-7}
\cline{9-11}
\cline{13-15}
4& 
10& 
12& 
& 
7& 
14& 
6& 
& 
9& 
21& 
41& 
& 
11& 
21& 
89 \\
\cline{1-3}
\cline{5-7}
\cline{9-11}
\cline{13-15}
4& 
12& 
11& 
& 
7& 
15& 
17& 
& 
9& 
22& 
14& 
& 
11& 
22& 
6 \\
\cline{1-3}
\cline{5-7}
\cline{9-11}
\cline{13-15}
4& 
15& 
4& 
& 
7& 
16& 
5& 
& 
9& 
24& 
13& 
& 
11& 
23& 
24 \\
\cline{1-3}
\cline{5-7}
\cline{9-11}
\cline{13-15}
4& 
16& 
2& 
& 
7& 
17& 
13& 
& 
9& 
25& 
9& 
& 
11& 
24& 
20 \\
\cline{1-3}
\cline{5-7}
\cline{9-11}
\cline{13-15}
4& 
18& 
1& 
& 
7& 
18& 
2& 
& 
9& 
27& 
3& 
& 
11& 
25& 
21 \\
\cline{1-3}
\cline{5-7}
\cline{9-11}
\cline{13-15}
5& 
2& 
11& 
& 
7& 
19& 
27& 
& 
9& 
28& 
7& 
& 
11& 
26& 
18 \\
\cline{1-3}
\cline{5-7}
\cline{9-11}
\cline{13-15}
5& 
3& 
43& 
& 
7& 
20& 
4& 
& 
9& 
29& 
19& 
& 
11& 
27& 
39 \\
\cline{1-3}
\cline{5-7}
\cline{9-11}
\cline{13-15}
5& 
4& 
39& 
& 
7& 
22& 
4& 
& 
9& 
30& 
2& 
& 
11& 
28& 
22 \\
\cline{1-3}
\cline{5-7}
\cline{9-11}
\cline{13-15}
5& 
6& 
4& 
& 
7& 
25& 
4& 
& 
9& 
31& 
8& 
& 
11& 
29& 
20 \\
\cline{1-3}
\cline{5-7}
\cline{9-11}
\cline{13-15}
5& 
7& 
5& 
& 
7& 
26& 
3& 
& 
9& 
32& 
4& 
& 
11& 
30& 
11 \\
\cline{1-3}
\cline{5-7}
\cline{9-11}
\cline{13-15}
5& 
8& 
18& 
& 
7& 
27& 
4& 
& 
9& 
33& 
2& 
& 
11& 
31& 
5 \\
\cline{1-3}
\cline{5-7}
\cline{9-11}
\cline{13-15}
5& 
9& 
10& 
& 
8& 
6& 
32& 
& 
9& 
34& 
4& 
& 
11& 
32& 
3 \\
\cline{1-3}
\cline{5-7}
\cline{9-11}
\cline{13-15}
5& 
10& 
34& 
& 
8& 
7& 
10& 
& 
10& 
7& 
7& 
& 
11& 
33& 
3 \\
\cline{1-3}
\cline{5-7}
\cline{9-11}
\cline{13-15}
5& 
11& 
23& 
& 
8& 
8& 
5& 
& 
10& 
8& 
14& 
& 
11& 
34& 
1 \\
\cline{1-3}
\cline{5-7}
\cline{9-11}
\cline{13-15}
5& 
12& 
19& 
& 
8& 
10& 
59& 
& 
10& 
12& 
17& 
& 
11& 
35& 
1 \\
\cline{1-3}
\cline{5-7}
\cline{9-11}
\cline{13-15}
5& 
13& 
5& 
& 
8& 
11& 
2& 
& 
10& 
13& 
3& 
& 
11& 
37& 
1 \\
\cline{1-3}
\cline{5-7}
\cline{9-11}
\cline{13-15}
5& 
15& 
12& 
& 
8& 
12& 
5& 
& 
10& 
14& 
4& 
& 
12& 
13& 
1 \\
\cline{1-3}
\cline{5-7}
\cline{9-11}
\cline{13-15}
5& 
16& 
6& 
& 
8& 
13& 
25& 
& 
10& 
15& 
20& 
& 
12& 
14& 
5 \\
\cline{1-3}
\cline{5-7}
\cline{9-11}
\cline{13-15}
5& 
19& 
1& 
& 
8& 
14& 
9& 
& 
10& 
16& 
30& 
& 
12& 
16& 
2 \\
\cline{1-3}
\cline{5-7}
\cline{9-11}
\cline{13-15}
5& 
20& 
1& 
& 
8& 
15& 
28& 
& 
10& 
17& 
18& 
& 
12& 
17& 
4 \\
\cline{1-3}
\cline{5-7}
\cline{9-11}
\cline{13-15}
6& 
3& 
11& 
& 
8& 
16& 
24& 
& 
10& 
18& 
36& 
& 
12& 
18& 
11 \\
\cline{1-3}
\cline{5-7}
\cline{9-11}
\cline{13-15}
6& 
4& 
76& 
& 
8& 
17& 
15& 
& 
10& 
19& 
83& 
& 
12& 
19& 
3 \\
\cline{1-3}
\cline{5-7}
\cline{9-11}
\cline{13-15}
6& 
6& 
7& 
& 
8& 
18& 
45& 
& 
10& 
20& 
78& 
& 
12& 
20& 
12 \\
\cline{1-3}
\cline{5-7}
\cline{9-11}
\cline{13-15}
\end{tabular}
\caption{There are ${\#}lib$ sub-libraries that can synthesize 
circuits with maximum length = $MaxLen$ and cost = $cost$.}
\label{tab12}
\end{center}
\end{table}

\begin{table}[htbp]
\small
\begin{center}
\begin{tabular}
{|p{15pt}|p{15pt}|p{25pt}|p{1pt}|p{15pt}|p{15pt}|p{25pt}|p{1pt}|p{15pt}|p{15pt}|p{25pt}|p{1pt}|p{15pt}|p{15pt}|p{25pt}|}
\cline{1-3}
\cline{5-7}
\cline{9-11}
\cline{13-15}
Max Len& 
cost& 
{\#}lib& 
& 
Max Len& 
cost& 
{\#}lib& 
& 
Max Len& 
cost& 
{\#}lib& 
& 
Max Len& 
cost& 
{\#}lib \\
\cline{1-3}
\cline{5-7}
\cline{9-11}
\cline{13-15}

12& 
21& 
52& 
& 
13& 
27& 
17& 
& 
14& 
28& 
27& 
& 
15& 
36& 
3 \\
\cline{1-3}
\cline{5-7}
\cline{9-11}
\cline{13-15}

12& 
22& 
30& 
& 
13& 
28& 
1& 
& 
14& 
29& 
3& 
& 
15& 
38& 
1 \\
\cline{1-3}
\cline{5-7}
\cline{9-11}
\cline{13-15}

12& 
23& 
17& 
& 
13& 
29& 
11& 
& 
14& 
30& 
3& 
& 
15& 
39& 
2 \\
\cline{1-3}
\cline{5-7}
\cline{9-11}
\cline{13-15}

12& 
24& 
52& 
& 
13& 
30& 
22& 
& 
14& 
31& 
2& 
& 
15& 
41& 
4 \\
\cline{1-3}
\cline{5-7}
\cline{9-11}
\cline{13-15}

12& 
25& 
44& 
& 
13& 
31& 
10& 
& 
14& 
32& 
9& 
& 
15& 
42& 
5 \\
\cline{1-3}
\cline{5-7}
\cline{9-11}
\cline{13-15}

12& 
26& 
42& 
& 
13& 
32& 
8& 
& 
14& 
34& 
8& 
& 
15& 
43& 
2 \\
\cline{1-3}
\cline{5-7}
\cline{9-11}
\cline{13-15}

12& 
27& 
14& 
& 
13& 
33& 
10& 
& 
14& 
35& 
9& 
& 
15& 
44& 
4 \\
\cline{1-3}
\cline{5-7}
\cline{9-11}
\cline{13-15}

12& 
28& 
23& 
& 
13& 
34& 
16& 
& 
14& 
36& 
11& 
& 
15& 
45& 
2 \\
\cline{1-3}
\cline{5-7}
\cline{9-11}
\cline{13-15}

12& 
29& 
31& 
& 
13& 
35& 
10& 
& 
14& 
37& 
6& 
& 
16& 
23& 
2 \\
\cline{1-3}
\cline{5-7}
\cline{9-11}
\cline{13-15}

12& 
30& 
28& 
& 
13& 
36& 
4& 
& 
14& 
38& 
5& 
& 
16& 
25& 
2 \\
\cline{1-3}
\cline{5-7}
\cline{9-11}
\cline{13-15}

12& 
31& 
6& 
& 
13& 
37& 
13& 
& 
14& 
39& 
1& 
& 
16& 
26& 
7 \\
\cline{1-3}
\cline{5-7}
\cline{9-11}
\cline{13-15}

12& 
32& 
12& 
& 
13& 
38& 
3& 
& 
14& 
40& 
7& 
& 
16& 
30& 
6 \\
\cline{1-3}
\cline{5-7}
\cline{9-11}
\cline{13-15}

12& 
33& 
13& 
& 
13& 
39& 
13& 
& 
14& 
41& 
1& 
& 
16& 
31& 
3 \\
\cline{1-3}
\cline{5-7}
\cline{9-11}
\cline{13-15}

12& 
34& 
5& 
& 
13& 
40& 
2& 
& 
14& 
42& 
2& 
& 
16& 
32& 
6 \\
\cline{1-3}
\cline{5-7}
\cline{9-11}
\cline{13-15}

12& 
35& 
9& 
& 
13& 
41& 
4& 
& 
14& 
45& 
1& 
& 
16& 
33& 
4 \\
\cline{1-3}
\cline{5-7}
\cline{9-11}
\cline{13-15}

12& 
36& 
4& 
& 
13& 
42& 
2& 
& 
15& 
21& 
3& 
& 
16& 
34& 
3 \\
\cline{1-3}
\cline{5-7}
\cline{9-11}
\cline{13-15}

12& 
37& 
4& 
& 
13& 
47& 
3& 
& 
15& 
23& 
4& 
& 
16& 
36& 
3 \\
\cline{1-3}
\cline{5-7}
\cline{9-11}
\cline{13-15}

12& 
38& 
9& 
& 
14& 
16& 
6& 
& 
15& 
24& 
4& 
& 
16& 
37& 
3 \\
\cline{1-3}
\cline{5-7}
\cline{9-11}
\cline{13-15}

13& 
16& 
1& 
& 
14& 
18& 
3& 
& 
15& 
25& 
15& 
& 
16& 
38& 
3 \\
\cline{1-3}
\cline{5-7}
\cline{9-11}
\cline{13-15}

13& 
18& 
5& 
& 
14& 
19& 
18& 
& 
15& 
26& 
8& 
& 
16& 
41& 
1 \\
\cline{1-3}
\cline{5-7}
\cline{9-11}
\cline{13-15}

13& 
19& 
63& 
& 
14& 
20& 
17& 
& 
15& 
28& 
16& 
& 
16& 
46& 
2 \\
\cline{1-3}
\cline{5-7}
\cline{9-11}
\cline{13-15}

13& 
20& 
8& 
& 
14& 
21& 
2& 
& 
15& 
29& 
10& 
& 
16& 
47& 
3 \\
\cline{1-3}
\cline{5-7}
\cline{9-11}
\cline{13-15}

13& 
21& 
15& 
& 
14& 
22& 
1& 
& 
15& 
30& 
1& 
& 
16& 
48& 
1 \\
\cline{1-3}
\cline{5-7}
\cline{9-11}
\cline{13-15}

13& 
22& 
17& 
& 
14& 
23& 
21& 
& 
15& 
31& 
9& 
& 
16& 
49& 
2 \\
\cline{1-3}
\cline{5-7}
\cline{9-11}
\cline{13-15}

13& 
23& 
36& 
& 
14& 
24& 
48& 
& 
15& 
32& 
16& 
& 
17& 
25& 
2 \\
\cline{1-3}
\cline{5-7}
\cline{9-11}
\cline{13-15}

13& 
24& 
10& 
& 
14& 
25& 
15& 
& 
15& 
33& 
9& 
& 
17& 
35& 
4 \\
\cline{1-3}
\cline{5-7}
\cline{9-11}
\cline{13-15}

13& 
25& 
55& 
& 
14& 
26& 
8& 
& 
15& 
34& 
1& 
& 
18& 
33& 
1 \\
\cline{1-3}
\cline{5-7}
\cline{9-11}
\cline{13-15}

13& 
26& 
19& 
& 
14& 
27& 
21& 
& 
15& 
35& 
1& 
& 
18& 
36& 
2 \\
\cline{1-3}
\cline{5-7}
\cline{9-11}
\cline{13-15}

\end{tabular}
\caption{Table \ref{tab12} cont.: There are ${\#}lib$ sub-libraries that can synthesize 
circuits with maximum length = $MaxLen$ and cost = $cost$.}
\label{tab13}
\end{center}
\end{table}

Table \ref{tab11} shows the maximum length circuits synthesized by a sub-library. There are 12 
sub-libraries that can synthesize a circuit with maximum length = 1. 
These are the sub-libraries that each of them 
contain a single generator from the main library shown in Eqn.\ref{12gatecycle}, where there are 
3 sub-libraries that synthesize a circuit with maximum length = 18. These three sub-libraries 
are $\{N_3,F_{32},F_{31},T_{123}\}$,$\{N_2,F_{23},F_{21},T_{132}\}$ and $\{N_1,F_{12},F_{13},T_{321}\}$. 
These sub-libraries are three examples from the 1960 universal reversible sub-libraries. 
Table \ref{tab12} continued in 
Table \ref{tab13} shows the same results as Table \ref{tab11} 
with more details on the cost of the circuits synthesized 
by these sub-libraries.

\begin{table}[htbp]
\small
\begin{center}
\begin{tabular}
{|p{15pt}|p{25pt}|p{1pt}|p{15pt}|p{25pt}|p{1pt}|p{15pt}|p{25pt}|p{1pt}|p{15pt}|p{25pt}|}
\cline{1-2}\cline{4-5}\cline{7-8}\cline{10-11}
Max cost&{\#}lib&&Max cost&{\#}lib&&Max cost& {\#}lib& & Max cost& {\#}lib \\\cline{1-2}\cline{4-5}\cline{7-8}\cline{10-11}
0& 7& & 14& 57& & 28& 227& & 42& 51 \\\cline{1-2}\cline{4-5}\cline{7-8}\cline{10-11}
1& 36& & 15& 75& & 29& 202& & 43& 25 \\\cline{1-2}\cline{4-5}\cline{7-8}\cline{10-11}
2& 69& & 16& 45& & 30& 131& & 44& 14 \\\cline{1-2}\cline{4-5}\cline{7-8}\cline{10-11}
3& 54& & 17& 23& & 31& 139& & 45& 7 \\\cline{1-2}\cline{4-5}\cline{7-8}\cline{10-11}
4& 138& & 18& 139& & 32& 229& & 46& 5 \\\cline{1-2}\cline{4-5}\cline{7-8}\cline{10-11}
5& 51& & 19& 153& & 33& 117& & 47& 10 \\\cline{1-2}\cline{4-5}\cline{7-8}\cline{10-11}
6& 98& & 20& 52& & 34& 95& & 48& 1 \\\cline{1-2}\cline{4-5}\cline{7-8}\cline{10-11}
7& 102& & 21& 76& & 35& 99& & 49& 3 \\\cline{1-2}\cline{4-5}\cline{7-8}\cline{10-11}
8& 94& & 22& 92& & 36& 91& & 50& 4 \\\cline{1-2}\cline{4-5}\cline{7-8}\cline{10-11}
9& 48& & 23& 56& & 37& 104& & 51& 3 \\\cline{1-2}\cline{4-5}\cline{7-8}\cline{10-11}
10& 31& & 24& 79& & 38& 69& & 52& 1 \\\cline{1-2}\cline{4-5}\cline{7-8}\cline{10-11}
11& 72& & 25& 180& & 39& 56& & 53& 2 \\\cline{1-2}\cline{4-5}\cline{7-8}\cline{10-11}
12& 56& & 26& 230& & 40& 37& & 54& 2 \\\cline{1-2}\cline{4-5}\cline{7-8}\cline{10-11}
13& 18& & 27& 207& & 41& 33& & &  \\\cline{1-2}\cline{4-5}\cline{7-8}\cline{10-11}
\end{tabular}
\caption{There are ${\#}lib$ libraries that can synthesize circuits with maximum 
cost =$Maxcost$.}
\label{tab14}
\end{center}
\end{table}

\begin{table}[htbp]
\small
\begin{center}
\begin{tabular}
{|p{15pt}|p{15pt}|p{25pt}|p{1pt}|p{15pt}|p{15pt}|p{25pt}|p{1pt}|p{15pt}|p{15pt}|p{25pt}|p{1pt}|p{15pt}|p{15pt}|p{25pt}|}
\cline{1-3}
\cline{5-7}
\cline{9-11}
\cline{13-15}
Max cost& 
Len& 
{\#}lib& 
& 
Max cost& 
Len& 
{\#}lib& 
& 
Max cost& 
Len& 
{\#}lib& 
& 
Max cost& 
Len& 
{\#}lib \\
\cline{1-3}  \cline{5-7} \cline{9-11} \cline{13-15} 
0& 
1& 
7& 
& 
15& 
7& 
22& 
& 
25& 
6& 
11& 
& 
31& 
12& 
12 \\
\cline{1-3}  \cline{5-7} \cline{9-11} \cline{13-15} 
1& 
1& 
36& 
& 
15& 
8& 
3& 
& 
25& 
7& 
21& 
& 
31& 
13& 
13 \\
\cline{1-3}  \cline{5-7} \cline{9-11} \cline{13-15} 
2& 
2& 
57& 
& 
15& 
9& 
2& 
& 
25& 
8& 
22& 
& 
32& 
8& 
1 \\
\cline{1-3}  \cline{5-7} \cline{9-11} \cline{13-15} 
2& 
4& 
12& 
& 
16& 
4& 
16& 
& 
25& 
9& 
93& 
& 
32& 
9& 
33 \\
\cline{1-3}  \cline{5-7} \cline{9-11} \cline{13-15} 
3& 
3& 
24& 
& 
16& 
5& 
12& 
& 
25& 
10& 
21& 
& 
32& 
10& 
36 \\
\cline{1-3}  \cline{5-7} \cline{9-11} \cline{13-15} 
3& 
4& 
1& 
& 
16& 
6& 
14& 
& 
25& 
11& 
12& 
& 
32& 
11& 
23 \\
\cline{1-3}  \cline{5-7} \cline{9-11} \cline{13-15} 
3& 
5& 
29& 
& 
16& 
7& 
3& 
& 
26& 
6& 
3& 
& 
32& 
12& 
89 \\
\cline{1-3}  \cline{5-7} \cline{9-11} \cline{13-15} 
4& 
4& 
129& 
& 
17& 
5& 
9& 
& 
26& 
7& 
20& 
& 
32& 
13& 
12 \\
\cline{1-3}  \cline{5-7} \cline{9-11} \cline{13-15} 
4& 
6& 
9& 
& 
17& 
6& 
2& 
& 
26& 
8& 
46& 
& 
32& 
14& 
34 \\
\cline{1-3}  \cline{5-7} \cline{9-11} \cline{13-15} 
5& 
1& 
6& 
& 
17& 
7& 
10& 
& 
26& 
9& 
51& 
& 
32& 
15& 
1 \\
\cline{1-3}  \cline{5-7} \cline{9-11} \cline{13-15} 
5& 
6& 
15& 
& 
17& 
8& 
2& 
& 
26& 
10& 
77& 
& 
33& 
9& 
18 \\
\cline{1-3}  \cline{5-7} \cline{9-11} \cline{13-15} 
5& 
7& 
30& 
& 
18& 
4& 
1& 
& 
26& 
11& 
25& 
& 
33& 
10& 
28 \\
\cline{1-3}  \cline{5-7} \cline{9-11} \cline{13-15} 
6& 
2& 
48& 
& 
18& 
5& 
3& 
& 
26& 
12& 
8& 
& 
33& 
11& 
27 \\
\cline{1-3}  \cline{5-7} \cline{9-11} \cline{13-15} 
6& 
6& 
38& 
& 
18& 
6& 
26& 
& 
27& 
7& 
20& 
& 
33& 
12& 
4 \\
\cline{1-3}  \cline{5-7} \cline{9-11} \cline{13-15} 
6& 
8& 
12& 
& 
18& 
7& 
73& 
& 
27& 
8& 
25& 
& 
33& 
13& 
3 \\
\cline{1-3}  \cline{5-7} \cline{9-11} \cline{13-15} 
7& 
3& 
54& 
& 
18& 
8& 
36& 
& 
27& 
9& 
47& 
& 
33& 
14& 
35 \\
\cline{1-3}  \cline{5-7} \cline{9-11} \cline{13-15} 
7& 
7& 
30& 
& 
19& 
5& 
4& 
& 
27& 
10& 
25& 
& 
33& 
15& 
2 \\
\cline{1-3}  \cline{5-7} \cline{9-11} \cline{13-15} 
7& 
8& 
4& 
& 
19& 
6& 
9& 
& 
27& 
11& 
73& 
& 
34& 
9& 
17 \\
\cline{1-3}  \cline{5-7} \cline{9-11} \cline{13-15} 
7& 
9& 
14& 
& 
19& 
7& 
127& 
& 
27& 
12& 
14& 
& 
34& 
10& 
33 \\
\cline{1-3}  \cline{5-7} \cline{9-11} \cline{13-15} 
8& 
4& 
18& 
& 
19& 
8& 
7& 
& 
27& 
13& 
3& 
& 
34& 
11& 
21 \\
\cline{1-3}  \cline{5-7} \cline{9-11} \cline{13-15} 
8& 
5& 
6& 
& 
19& 
9& 
6& 
& 
28& 
8& 
22& 
& 
34& 
12& 
5 \\
\cline{1-3}  \cline{5-7} \cline{9-11} \cline{13-15} 
8& 
6& 
12& 
& 
20& 
5& 
5& 
& 
28& 
9& 
53& 
& 
34& 
13& 
6 \\
\cline{1-3}  \cline{5-7} \cline{9-11} \cline{13-15} 
8& 
8& 
34& 
& 
20& 
6& 
12& 
& 
28& 
10& 
48& 
& 
34& 
14& 
4 \\
\cline{1-3}  \cline{5-7} \cline{9-11} \cline{13-15} 
8& 
9& 
19& 
& 
20& 
7& 
22& 
& 
28& 
11& 
15& 
& 
34& 
15& 
6 \\
\cline{1-3}  \cline{5-7} \cline{9-11} \cline{13-15} 
8& 
10& 
5& 
& 
20& 
8& 
5& 
& 
28& 
12& 
55& 
& 
34& 
16& 
3 \\
\cline{1-3}  \cline{5-7} \cline{9-11} \cline{13-15} 
9& 
5& 
42& 
& 
20& 
9& 
2& 
& 
28& 
13& 
33& 
& 
35& 
9& 
10 \\
\cline{1-3}  \cline{5-7} \cline{9-11} \cline{13-15} 
9& 
10& 
6& 
& 
20& 
10& 
6& 
& 
28& 
14& 
1& 
& 
35& 
10& 
26 \\
\cline{1-3}  \cline{5-7} \cline{9-11} \cline{13-15} 
10& 
3& 
4& 
& 
21& 
6& 
19& 
& 
29& 
7& 
6& 
& 
35& 
11& 
41 \\
\cline{1-3}  \cline{5-7} \cline{9-11} \cline{13-15} 
10& 
4& 
27& 
& 
21& 
7& 
27& 
& 
29& 
8& 
11& 
& 
35& 
12& 
13 \\
\cline{1-3}  \cline{5-7} \cline{9-11} \cline{13-15} 
11& 
3& 
17& 
& 
21& 
8& 
12& 
& 
29& 
9& 
85& 
& 
35& 
13& 
6 \\
\cline{1-3}  \cline{5-7} \cline{9-11} \cline{13-15} 
11& 
4& 
16& 
& 
21& 
10& 
6& 
& 
29& 
10& 
41& 
& 
35& 
16& 
3 \\
\cline{1-3}  \cline{5-7} \cline{9-11} \cline{13-15} 
11& 
5& 
39& 
& 
21& 
11& 
12& 
& 
29& 
11& 
44& 
& 
36& 
9& 
3 \\
\cline{1-3}  \cline{5-7} \cline{9-11} \cline{13-15} 
12& 
4& 
19& 
& 
22& 
6& 
2& 
& 
29& 
12& 
8& 
& 
36& 
10& 
18 \\
\cline{1-3}  \cline{5-7} \cline{9-11} \cline{13-15} 
12& 
5& 
8& 
& 
22& 
7& 
45& 
& 
29& 
14& 
7& 
& 
36& 
11& 
18 \\
\cline{1-3}  \cline{5-7} \cline{9-11} \cline{13-15} 
12& 
6& 
29& 
& 
22& 
8& 
45& 
& 
30& 
8& 
11& 
& 
36& 
12& 
19 \\
\cline{1-3}  \cline{5-7} \cline{9-11} \cline{13-15} 
13& 
5& 
9& 
& 
23& 
6& 
2& 
& 
30& 
9& 
24& 
& 
36& 
13& 
13 \\
\cline{1-3}  \cline{5-7} \cline{9-11} \cline{13-15} 
13& 
6& 
9& 
& 
23& 
7& 
13& 
& 
30& 
10& 
47& 
& 
36& 
14& 
20 \\
\cline{1-3}  \cline{5-7} \cline{9-11} \cline{13-15} 
14& 
3& 
2& 
& 
23& 
8& 
23& 
& 
30& 
11& 
26& 
& 
37& 
10& 
25 \\
\cline{1-3}  \cline{5-7} \cline{9-11} \cline{13-15} 
14& 
4& 
34& 
& 
23& 
9& 
15& 
& 
30& 
12& 
22& 
& 
37& 
11& 
24 \\
\cline{1-3}  \cline{5-7} \cline{9-11} \cline{13-15} 
14& 
6& 
12& 
& 
23& 
10& 
3& 
& 
30& 
15& 
1& 
& 
37& 
12& 
20 \\
\cline{1-3}  \cline{5-7} \cline{9-11} \cline{13-15} 
14& 
7& 
9& 
& 
24& 
6& 
4& 
& 
31& 
7& 
3& 
& 
37& 
13& 
16 \\
\cline{1-3}  \cline{5-7} \cline{9-11} \cline{13-15} 
15& 
3& 
1& 
& 
24& 
7& 
5& 
& 
31& 
8& 
1& 
& 
37& 
14& 
17 \\
\cline{1-3}  \cline{5-7} \cline{9-11} \cline{13-15} 
15& 
4& 
37& 
& 
24& 
8& 
43& 
& 
31& 
9& 
31& 
& 
37& 
15& 
2 \\
\cline{1-3}  \cline{5-7} \cline{9-11} \cline{13-15} 
15& 
5& 
8& 
& 
24& 
9& 
19& 
& 
31& 
10& 
33& 
& 
38& 
10& 
9 \\
\cline{1-3}  \cline{5-7} \cline{9-11} \cline{13-15} 
15& 
6& 
2& 
& 
24& 
10& 
8& 
& 
31& 
11& 
46& 
& 
38& 
11& 
23 \\
\cline{1-3}  \cline{5-7} \cline{9-11} \cline{13-15} 
\end{tabular}
\caption{There are ${\#}lib$ libraries that can synthesize circuits with maximum 
cost = $Maxcost$ and length = $Len$.}
\label{tab15}
\end{center}
\end{table}

\begin{table}[htbp]
\small
\begin{center}
\begin{tabular}
{|p{15pt}|p{15pt}|p{25pt}|p{1pt}|p{15pt}|p{15pt}|p{25pt}|p{1pt}|p{15pt}|p{15pt}|p{25pt}|p{1pt}|p{15pt}|p{15pt}|p{15pt}|}
\cline{1-3}  \cline{5-7} \cline{9-11} \cline{13-15} 
Max cost& 
Len& 
{\#}lib& 
& 
Max cost& 
Len& 
{\#}lib& 
& 
Max cost& 
Len& 
{\#}lib& 
& 
Max cost& 
Len& 
{\#}lib \\
\cline{1-3}  \cline{5-7} \cline{9-11} \cline{13-15} 
38& 
12& 
13& 
& 
40& 
14& 
3& 
& 
43& 
13& 
9& 
& 
47& 
14& 
5 \\
\cline{1-3}  \cline{5-7} \cline{9-11} \cline{13-15} 
38& 
13& 
4& 
& 
40& 
18& 
3& 
& 
43& 
14& 
4& 
& 
48& 
13& 
1 \\
\cline{1-3}  \cline{5-7} \cline{9-11} \cline{13-15} 
38& 
14& 
3& 
& 
41& 
11& 
2& 
& 
44& 
11& 
1& 
& 
49& 
14& 
3 \\
\cline{1-3}  \cline{5-7} \cline{9-11} \cline{13-15} 
38& 
15& 
14& 
& 
41& 
12& 
12& 
& 
44& 
13& 
10& 
& 
50& 
13& 
1 \\
\cline{1-3}  \cline{5-7} \cline{9-11} \cline{13-15} 
38& 
16& 
3& 
& 
41& 
13& 
13& 
& 
44& 
14& 
1& 
& 
50& 
15& 
3 \\
\cline{1-3}  \cline{5-7} \cline{9-11} \cline{13-15} 
39& 
10& 
1& 
& 
41& 
14& 
6& 
& 
44& 
15& 
2& 
& 
51& 
14& 
2 \\
\cline{1-3}  \cline{5-7} \cline{9-11} \cline{13-15} 
39& 
11& 
15& 
& 
42& 
11& 
19& 
& 
45& 
13& 
4& 
& 
51& 
15& 
1 \\
\cline{1-3}  \cline{5-7} \cline{9-11} \cline{13-15} 
39& 
12& 
27& 
& 
42& 
12& 
18& 
& 
45& 
14& 
3& 
& 
52& 
15& 
1 \\
\cline{1-3}  \cline{5-7} \cline{9-11} \cline{13-15} 
39& 
13& 
13& 
& 
42& 
13& 
10& 
& 
46& 
12& 
1& 
& 
53& 
15& 
2 \\
\cline{1-3}  \cline{5-7} \cline{9-11} \cline{13-15} 
40& 
10& 
6& 
& 
42& 
14& 
2& 
& 
46& 
14& 
2& 
& 
54& 
15& 
2 \\
\cline{1-3}  \cline{5-7} \cline{9-11} 
40& 
11& 
9& 
& 
42& 
15& 
2& 
& 
46& 
15& 
2& 
& 
& 
& 
 \\
\cline{1-3}  \cline{5-7} \cline{9-11} 
40& 
12& 
14& 
& 
43& 
11& 
6& 
& 
47& 
12& 
2& 
& 
& 
& 
 \\
\cline{1-3}  \cline{5-7} \cline{9-11} 
40& 
13& 
2& 
& 
43& 
12& 
6& 
& 
47& 
13& 
3& 
& 
& 
& 
 \\
\cline{1-3}  \cline{5-7} \cline{9-11} \cline{13-15} 
\end{tabular}
\caption{Table \ref{tab15} cont.: There are ${\#}lib$ libraries that can synthesize circuits with maximum 
cost = $Maxcost$ and length = $Len$.}
\label{tab16}
\end{center}
\end{table}

Table \ref{tab14} shows the maximum cost circuits synthesized by a sub-library. There are 7 
sub-libraries that can synthesize a circuit with maximum cost = 0. 
These are the sub-libraries that each of them contain a combination of $N$ generators, where there are 
2 sub-libraries that synthesize a circuit with maximum cost = 54. These two sub-libraries 
are $\{N_1,N_3,F_{32},T_{123},T_{321}\}$ and $\{N_1,F_{12},T_{123},T_{321}\}$. These are another 
two examples from the 1960 universal reversible sub-libraries. 
Table \ref{tab15} continued in Table \ref{tab16} shows the same results as Table \ref{tab14} with more 
details on the cost of the circuits synthesized by these sub-libraries.

The above results show that there are 1960 sub-libraries, each can be used as a universal reversible gate 
library:
\begin{enumerate}
\item The sub-library that synthesize the best maximum length circuits is the main library,

\[
\{N_1,N_2,N_3,F_{12},F_{13},F_{23},F_{21},F_{32},F_{31},T_{123},T_{132},T_{321}\},
\]

\noindent where the best maximum length = 8 gates with cost = 20 
and maximum cost = 23 with circuit length =7. 

\item
The sub-library that synthesize the best maximum cost circuit is,
 
\[
\{N_1,N_2,N_3,F_{12},F_{13},F_{23},F_{21},F_{32},F_{31},T_{132},T_{321}\},
\] 

\noindent where the best maximum cost = 22 with circuit length =7 
and the maximum length = 9 gates with cost = 11. 

\item
The sub-library that synthesize the worst maximum length circuits is,
  
\[
\{N_3,F_{32},F_{31},T_{123}\},
\] 

\noindent where the worst maximum length = 18 gates with cost = 36 
and maximum cost = 40 with circuit length =18.

\item 
The sub-library that synthesize the worst maximum cost circuit is,

\[
\{N_1,F_{12},T_{123},T_{321}\},
\] 

\noindent where the worst maximum cost = 54 with circuit length =15
and the maximum length = 16 gates with cost = 49. 
\end{enumerate}

Non of the 1960 universal reversible gate libraries can synthesize a circuit with the best 
maximum cost which is 17 as shown in Table \ref{tab8}. This best maximum cost comes from sub-libraries 
that are not universal, for example, the sub-library,
\[
\{N_1,N_2,F_{13},F_{23},F_{31},T_{123},T_{321}\},
\] 

\noindent is not universal since it can syntheise circuits for 1151 specifications only, 
where its best maximum cost = 17 with circuit length =7 and the maximum length = 7 gates with cost = 8.

\section{Conclusion}

By reducing the representation of the reversible circuit synthesis problem to permutation group,  
Schreier-Sims Algorithm for the strong generating set-finding problem is used to put tight bounds 
on the synthesis of 3-bit reversible circuits using the NFT library. Using group-theory algebraic 
software GAP shows that,

\begin{enumerate}

\item The minimum length of a circuit ranges from 1 to 8 gate(s) with average length = 5.865.

\item The maximum length of a circuit ranges from 7 to 18 gates with average length = 14.639.

\item The minimum cost of a reversible circuit ranges from 1 to 17 with average cost =  11.769.
 
\item The maximum cost of a reversible circuit ranges from 18 to 54 with average cost =  40.759.

\end{enumerate}
 
The analysis shows that there 1960 universal reversible sub-libraries from the NFT library. 
The upper and lower bounds on the length of the circuits come from using the universal reversible sub-libraries while 
the upper bound on the cost of the circuits comes from using the universal reversible sub-libraries, while 
the lower bounds on the cost of the circuits comes from other sub-libraries which are not necessary universal.

The same sort of analysis is applied to other libraries such as NFP, NFFr, NFPT, NFTFr and NFPFr. 
The tight bounds for these libraries are under preparation.

\end{document}